\documentstyle[prd,eqsecnum,aps]{revtex}
\def\be{\begin{equation}}
\def\ee{\end{equation}}
\def\bea{\begin{eqnarray}}
\def\eea{\end{eqnarray}}
\def\l{\label}
\def\ct{\cite}
\def\r{\ref}
\def\lgth{[\,\mbox{length}\,]}
\def\gam{\gamma}
\def\d{\delta}
\def\eps{\epsilon}
\def\gam{\gamma}
\def\Th{\Theta}
\def\sig{\sigma}
\def\om{\omega}
\def\udot{\dot{u}}
\def\D{\mbox{D}}
\def\p{{\bf e}}
\def\ptl{\partial}
\def\lgl{\langle}
\def\rgl{\rangle}
\def\tl{\tilde}
\def\hsp5{\hspace{5mm}}
\newcommand{\sfrac}[2]{{\textstyle{#1\over#2}}}
\def\case#1/#2{\textstyle\frac{#1}{#2}}
\def\cqg{{Class. Quantum Grav.\/} }
\def\grg{{Gen. Rel. Grav.\/} }
\def\prd{{Phys. Rev.\/} D }
\def\prl{{Phys. Rev. Lett.\/} }
\newcommand{\enl}{\\\hfill\rule{0pt}{0pt}}
\begin{document}
\preprint{UCT--98003}
\draft
\title{Causal propagation of geometrical fields in relativistic
cosmology}
\author{Henk van Elst\thanks{Electronic address:
henk@gmunu.mth.uct.ac.za} and
George F. R. Ellis\thanks{Electronic address:
ellis@maths.uct.ac.za}}

\address{Department of Mathematics and Applied
Mathematics, University of Cape Town\\
Rondebosch 7701, Cape Town, South Africa}

\date{August 30, 1998}
\maketitle
\begin{abstract}
We employ the extended $1+3$ orthonormal frame formalism for fluid
spacetime geometries $\left(\,{\cal M},\,{\bf g},\,{\bf
u}\,\right)$, which contains the Bianchi field equations for the
Weyl curvature, to derive a 44-D evolution system of first-order
symmetric hyperbolic form for a set of geometrically defined
dynamical field variables. Describing the matter source fields
phenomenologically in terms of a barotropic perfect fluid, the
propagation velocities $v$ (with respect to matter-comoving
observers that Fermi-propagate their spatial reference frames) of
disturbances in the matter and the gravitational field, represented
as wavefronts by the characteristic 3-surfaces of the system, are
obtained. In particular, the Weyl curvature is found to account for
two (non-Lorentz-invariant) Coulomb-like characteristic eigenfields
propagating with $v = 0$ and four transverse characteristic
eigenfields propagating with $|v| = 1$, which are well known, and
four (non-Lorentz-invariant) longitudinal characteristic
eigenfields propagating with $|v| = \sfrac{1}{2}$. The implications
of this result are discussed in some detail and a parallel is drawn
to the propagation of irregularities in the matter distribution. In
a worked example, we specialise the equations to cosmological
models in locally rotationally symmetric class II and include the
constraints into the set of causally propagating dynamical
variables.

\end{abstract}
\widetext
\section{Introduction}
The natural way to relate gravitational fields to physical
measurements is by detecting their tidal and their volume-changing
effects: monitoring the relative accelerations between a set of
test particles in free fall, measuring the gravitationally induced
deformations between the ends of a semi-rigid detector, or
detecting the distortions of the cross section of an infinitesimal
beam of light that travels freely through spacetime. This viewpoint
has been emphasised strongly by a number of authors, particularly,
Pirani \ct{pir56}, Szekeres \ct{sze65}, and Wheeler \ct{whe90}. In
mathematical terms, the connection between a gravitational field
and the measurement process is made through the geodesic deviation
equation (`GDE'), introduced into general relativity by Synge
\ct{syn34}, wherein the Riemann curvature of a given spacetime
manifold $\left(\,{\cal M},\,{\bf g}\,\right)$ provides the force
term responsible for the worldlines of test matter to deviate from
the straight lines of Minkowskian geometry.\footnote{Or through the
closely associated relative deviation equation for particles
subject to non-gravitational forces, and, hence, moving
non-geodesically (\,see, e.g., Ehlers \ct{ehl61}\,); however, the
essential gravitational effects are already incorporated in the
GDE, so we concentrate on that equation in this paper.} If the
(timelike or null) tangents to these worldlines are denoted by
$V^{a}$, the second rate of change along $V^{a}$ of the
(orthogonal) connecting vectors $\eta^{a}$ is determined by the
Riemann curvature (\,see, e.g., Ref. \ct{wal84}\,):
\be
\l{gde}
V^{c}\nabla_{c}(V^{b}\nabla_{b}\eta^{a}) 
= - \,R^{a}{}_{bcd}\, V^{b}\,\eta^{c}\,V^{d} \ .
\ee
Geometrically the Riemann curvature is often defined as the linear
operator which encodes the non-commutativity of two covariant
derivatives on $\left(\,{\cal M},\,{\bf g}\,\right)$, i.e.
\be
\l{ricid}
2\,\nabla_{[a}\nabla_{b]}v^{c} := R_{ab}{}^{c}{}_{d}\,v^{d} \ ,
\ee
for any vector $v^{a}$, from which the above equation follows.  As
is standard, the Riemann curvature tensor can be decomposed into
the Weyl curvature tensor, the Ricci curvature tensor, and the
Ricci curvature scalar according to
\be
\l{riem}
R_{abcd} = C_{abcd} + R_{a[c}\,g_{d]b} - R_{b[c}\,g_{d]a}
- \sfrac{1}{3}\,R\,g_{a[c}\,g_{d]b} \ .
\ee
Consequently these are the force terms in the GDE. If, as is the
case in relativistic cosmology, there exists an invariantly defined
future-directed unit timelike vector field ${\bf u}$, the Weyl and
the Ricci curvatures of $\left(\,{\cal M},\,{\bf g},\,{\bf
u}\,\right)$ can be further decomposed with respect to the group of
spatial rotations in a $1+3$ covariant way \ct{ehl61,ell71}
according to
\bea
\l{weyl}
C^{ab}{}_{cd}({\bf u}) & = & 4\,u^{[a}\,u_{[c}\,E^{b]}{}_{d]}
+ 4\,h^{[a}{}_{[c}\,E^{b]}{}_{d]}
+ 2\,\eps^{abe}\,u_{[c}\,H_{d]e} + 2\,\eps_{cde}\,u^{[a}\,H^{b]e}
\\ \nonumber \\
\l{ric}
R_{ab}({\bf u}) & = & \sfrac{1}{2}\,(\mu+3p-2\Lambda)\,u_{a}\,u_{b}
+ 2\,q_{(a}\,u_{b)} + \sfrac{1}{2}\,(\mu-p+2\Lambda)\,h_{ab}
+ \pi_{ab} \ ,
\eea
respectively, where in Eq. (\r{weyl}) $E_{ab}$ and $H_{ab}$ are the
electric and magnetic Weyl curvature tensors (relative to ${\bf
u}$), and in Eq. (\r{ric}) the Einstein field equations (`EFE')
$R_{ab} - \sfrac{1}{2}\,R\,g_{ab} + \Lambda\,g_{ab} = T_{ab}$ have
been used to equate the Ricci curvature to the
energy-momentum-stress of the matter source fields, $T_{ab}({\bf
u}) := \mu\,u_{a}\,u_{b} + 2\,q_{(a}\,u_{b)} + p\,h_{ab} +
\pi_{ab}$. The matter fluid variables $\mu$, $q^{a}$, $p$, and
$\pi_{ab}$ are the energy density, energy current density,
isotropic pressure, and tracefree anisotropic pressure (relative to
${\bf u}$); $h_{ab} := g_{ab} + u_{a}\,u_{b}$ is the projection
tensor into the instantaneous rest 3-spaces orthogonal to ${\bf u}$
with 3-volume element $\eps_{abc} :=
-\,\eps_{defg}\,h^{d}\!_{a}\,h^{e}\!_{b}\,h^{f}\!_{c} \,u^{g}$.

It is clear, then, that all these terms enter into the GDE and can
have measurable physical effects; hence, their propagation
behaviours are of physical importance. Tying the $1+3$ viewpoint to
an orthonormal frame description which includes the full Riemann
curvature, the dynamical equations for the associated geometrical
field variables are provided by the frame basis commutation
relations, the Jacobi, the Ricci and also the Bianchi field
equations \ct{hveugg97}. These determine the propagation modes of
all the above quantities, but not immediately in a mathematically
most desirable way, because the hyperbolic nature of the overall
equation system is not clearly apparent.

Recently research activity in the arena of numerical relativity
(\,see, e.g.,
Refs. \ct{bonetal95,abretal97,putear96,estetal97,reu98}\,) has
triggered a growing interest in formulating evolution systems for
dynamical field variables in such a way that their propagation
velocities transparently respect relativistic causality, and where
physically motivated boundary conditions can be implemented more
easily than in the standard ADM $3+1$ formalism \ct{adm62}. A
natural framework for this purpose is provided by the first-order
symmetric hyperbolic (`FOSH') form for sets of (quasi-linear)
partial differential equations, that Friedrichs introduced in the
mid-fifties \ct{fri54}. Fortunately, most sets of dynamical
equations occurring in theoretical physics can be cast into such a
form \ct{ger96}.\footnote{Incidentally, Geroch \ct{ger96} leaves
the case of the (vacuum) EFE in his nice work open; as other papers
cited in this introduction and the present work show, a FOSH
evolution system for general relativity can be obtained, e.g., by
either employing the Bianchi field equations or introducing
auxiliary variables based on first derivatives of the metric.} For
a physicist this is especially attractive for at least two reasons:
first, there do exist mathematical proofs that show (local)
existence and uniqueness of solutions to the related Cauchy initial
value problem; secondly, FOSH evolution systems allow for the
propagation of data that are {\em non-analytic\/} across certain
(characteristic) 3-surfaces, which can be interpreted as
wavefronts. A standard reference on both aspects is the book by
Courant and Hilbert \ct{couhil62}.

Suppose a collection of dependent field variables $u^{A} =
u^{A}(x^{\mu})$, which are functions of a set of local spacetime
coordinates $\{\,x^{\mu}\,\}$, satisfies the equation system
\be
\l{fosh}
M^{AB\,\mu}\,\ptl_{\mu} u_{B} = N^{A} \ .
\ee
The objects $M^{AB\,\mu} = M^{AB\,\mu}(x^{\nu},u^{C})$ and $N^{A} =
N^{A}(x^{\mu},u^{B})$ denote four {\em symmetric\/} matrices and a
vector, respectively, each acting in a space of dimension equal to
the number of dependent fields. Conventionally one refers to the
left-hand side of Eq. (\r{fosh}) --- the `principle part' --- as
representing dynamical interactions between certain physical
fields, the right-hand side representing kinematical ones
(\,cf. Geroch \ct{ger96}\,). The set of equations (\r{fosh}) is
{\em hyperbolic\/} if the contraction $M^{AB\,\mu}\,n_{\mu}$ with
the coordinate components of some past-directed timelike 1-form
$n_{a}$ yields a positive-definite matrix; it is {\em causal\/} if
this contraction is positive-definite for {\em all\/} past-directed
timelike 1-forms $n_{a}$ \ct{gerlin90}. If it satisfies all these
conditions, it is said to be a FOSH evolution system.

In the present work we inquire into the optimal formulation of a
FOSH evolution system for cosmological models $\left(\,{\cal
M},\,{\bf g},\,{\bf u}\,\right)$ with barotropic perfect fluid
matter source fields, starting from the dynamical equations of the
extended $1+3$ orthonormal frame (`ONF') formalism which includes
the Weyl curvature variables, as discussed recently in
Ref. \ct{hveugg97}. The fact that the Bianchi field equations for
the Weyl curvature, which are of {\em third\/} order in the
derivatives of the metric ${\bf g}$, naturally lead to FOSH
equations, was already recognised by Friedrich in the early
eighties \ct{fri81}. Moreover, lately he employed them in work on
the derivation of FOSH evolution systems for vacuum gravitational
fields \ct{fri96} and for gravitating perfect fluid bodies
\ct{fri98}. Our choice of dependent field variables as provided by
the extended $1+3$ ONF formalism shares the same philosophy as
Friedrich's, but additionally we have fully decomposed the
spacetime connection into its irreducible parts (hence, obtaining
more transparent equations), and we introduce a set of local
coordinates by following the $1+3$ threading picture
\ct{janetal92,boedra95}. The latter is of particular relevance when
the timelike reference congruence of the setting has {\em
non-zero\/} vorticity.

Special attention in our investigation will be given to the
properties of a longitudinal component in the Weyl curvature as
described by Szekeres in work on the propagation of gravitational
fields in matter \ct{sze65,sze66}. Friedrich (in the vacuum case)
discounted the propagation modes for these components as not being
physically significant, because the characteristic 3-surfaces along
which they travel are not Lorentz-invariant \ct{fri96}. We suggest
an opposite view: that despite this feature, these modes must be
seen as physically significant because of their detectable effects
as determined by the GDE. We briefly consider some consequences of
this suggestion.

We specify our conventions, choice of units, and some of our
notation in appendix \r{app1}.

\section{General $1+3$-decomposed dynamical equations}
\l{sec:13eqs}
For easy reference, we dedicate this section to a complete
exposition of the $1+3$ ONF dynamical equations, where the
energy-momentum-stress of the matter source fields is assumed to
have the form of a general dissipative relativistic fluid. As
described in detail in Ref. \ct{hveugg97}, for a cosmological model
$\left(\,{\cal M},\,{\bf g},\,{\bf u}\,\right)$ a set of four
orthogonal unit basis fields $\{\,\p_{a}\,\}$ is chosen and its
timelike member $\p_{0}$ is identified with the unit tangent ${\bf
u}$ to the matter flow lines (here as viewed from the `particle
frame' perspective). This fixing implies that we reduce the initial
six-parameter Lorentz transformation freedom in the choice of
$\{\,\p_{a}\,\}$ to a three-parameter freedom that allows for
arbitrary rotations of the spatial frame $\{\,\p_{\alpha}\,\}$
only.

\subsection{Connection components and commutators}
The 24 algebraically independent frame components of the spacetime
connection $\Gamma^{a}{}_{bc}$ can be split into the
set\footnote{To facilitate comparison, we indicate how these frame
components of the connection relate to the respective variables
used by Friedrich in Ref. \ct{fri98}, which we denote by
$\Gamma_{\!F\,abc}$.}
\bea
\l{onfgam1}
\Gamma_{\alpha 00} & = & \udot_{\alpha} \ = \ \Gamma_{\!F\,00\alpha}
\\
\Gamma_{\alpha 0\beta} & = & \sfrac{1}{3}\,\Th\,\d_{\alpha\beta}
+ \sig_{\alpha\beta} - \eps_{\alpha\beta\gam}\,\om^{\gam}
\ = \ \Gamma_{\!F\,0\beta\alpha} \\
\Gamma_{\alpha\beta 0} & = & \eps_{\alpha\beta\gam}\,\Omega^{\gam}
\ = \ \Gamma_{\!F\,0\alpha\beta} \\
\l{onfgam4}
\Gamma_{\alpha\beta\gam} & = & 2\,a_{[\alpha}\,\d_{\beta]\gam}
+ \eps_{\gam\delta[\alpha}\,n^{\delta}\!_{\beta]}
+ \sfrac{1}{2}\,\eps_{\alpha\beta\delta}\,n^{\delta}\!_{\gam}
\ = \ \Gamma_{\!F\,\beta\gam\alpha} \ .
\eea
It contains the familiar kinematical fluid variables:
$\udot^{\alpha}$, its relativistic acceleration, $\Th$, its volume
rate of expansion, $\sig_{\alpha\beta} = \sig_{\beta\alpha}$, its
rate of shear (with $\sig^{\alpha}\!_{\alpha} = 0$), and
$\om^{\alpha}$, its vorticity. Moreover, $\Omega^{\alpha}$ is the
rate of rotation of the spatial frame $\{\,\p_{\alpha}\,\}$ with
respect to a Fermi-propagated basis. Finally, $a^{\alpha}$ and
$n_{\alpha\beta} = n_{\beta\alpha}$ are 9 spatial commutation
functions.

With our definition of the covariant derivative on $\left(\,{\cal
M},\,{\bf g},\,{\bf u}\,\right)$, given in appendix \r{app1}, and
Eqs. (\r{onfgam1}) - (\r{onfgam4}), the condition that the
spacetime connection be torsion-free, $\nabla_{[a}\nabla_{b]}f = 0$
for any spacetime scalar $f$, translates into the commutator
equations
\bea
\l{onfcomts}
[\,\p_{0}, \p_{\alpha}\,]\,(f) & = &
\udot_{\alpha}\ \p_{0}(f) - [\ \sfrac{1}{3}\,\Th\,
\d_{\alpha}{}^{\beta} + \sig_{\alpha}{}^{\beta}
+ \eps_{\alpha}{}^{\beta}{}_{\gam}\,(\om^{\gam}
+\Omega^{\gam})\ ]\ \p_{\beta}(f)
\\ \nonumber \\
\l{onfcomss}
[\,\p_{\alpha}, \p_{\beta}\,]\,(f) & = &
2\,\eps_{\alpha\beta\gam}\,\om^{\gam}\ \p_{0}(f)
+ [\ 2\,a_{[\alpha}\,\d^{\gam}\!_{\beta]}
+ \eps_{\alpha\beta\delta}\,n^{\delta\gam}\ ]\ \p_{\gam}(f) \ .
\eea
%
\subsection{Constraints}
The following set of relations does not contain any frame
derivatives with respect to $\p_{0}$. Hence, it is commonplace to
refer to these relations as `constraints'.\footnote{Even though
this terminology is problematic in the generic case when $\p_{0} =
{\bf u}$ has {\em non-zero\/} vorticity and local coordinates are
introduced according to Eq. (\r{onf13}) below.} From the Ricci
identity for ${\bf u}$ as well as the Jacobi identity we have the
$\D_{b}\sig^{ab}$-equation $(C_{1})^{\alpha}$, which, in
Hamiltonian treatments of the EFE, is also referred to as the
`momentum constraint', the $\D_{a}\om^{a}$-equation $(C_{2})$, and
the $H_{ab}$-constraint $(C_{3})^{\alpha\beta}$, respectively; the
once-contracted second Bianchi identity yields the $\D_{b}E^{ab}$-
and $\D_{b}H^{ab}$-equations $(C_{4})^{\alpha}$ and
$(C_{5})^{\alpha}$ \ct{ell71,hve96}; the constraint
$(C_{J})^{\alpha}$ again arises from the Jacobi identity, while,
finally, $(C_{G})^{\alpha\beta}$ and $(C_{G})$ stem from the
EFE. In detail,
\bea
\l{onfdivsig}
0 & = & (C_{1})^{\alpha} \ := \ (\p_{\beta} - 3\,a_{\beta})\,
(\sig^{\alpha\beta}) - \sfrac{2}{3}\,\d^{\alpha\beta}\,
\p_{\beta}(\Th) - n^{\alpha}\!_{\beta}\,\om^{\beta} + q^{\alpha}
\nonumber \\
& & \hspace{25mm} + \ \epsilon^{\alpha\beta\gamma}\,
[\ (\p_{\beta} + 2\,\udot_{\beta} - a_{\beta})\,
(\om_{\gamma}) - n_{\beta\delta}\,\sig^{\delta}\!_{\gamma}\ ]
\\ \nonumber \\
\l{onfdivom}
0 & = & (C_{2}) \ := \ (\p_{\alpha} - \udot_{\alpha}
- 2\,a_{\alpha})\,(\om^{\alpha})
\\ \nonumber \\
\l{onfhconstr}
0 & = & (C_{3})^{\alpha\beta} \ := \ H^{\alpha\beta}
+ (\d^{\gam\lgl\alpha}\,\p_{\gam} + 2\,\udot^{\lgl\alpha}
+ a^{\lgl\alpha})\,(\om^{\beta\rgl}) - \sfrac{1}{2}\,
n^{\gam}\!_{\gam}\,\sig^{\alpha\beta}
+ 3\,n^{\lgl\alpha}\!_{\gam}\,\sig^{\beta\rgl\gamma} \nonumber \\
& & \hspace{25mm} - \ \eps^{\gam\delta\lgl\alpha}\,[\ (\p_{\gam}
- a_{\gam})\,(\sig^{\beta\rgl}\!_{\delta})
+ n^{\beta\rgl}\!_{\gam}\,\om_{\delta}\ ]
\\ \nonumber \\
\l{onfdive}
0 & = & (C_{4})^{\alpha} \ := \ (\p_{\beta} - 3\,a_{\beta})
\,(E^{\alpha\beta}+\sfrac{1}{2}\,\pi^{\alpha\beta})
- \sfrac{1}{3}\,\d^{\alpha\beta}\,\p_{\beta}(\mu)
+ \sfrac{1}{3}\,\Th\,q^{\alpha}
- \sfrac{1}{2}\,\sig^{\alpha}\!_{\beta}\,q^{\beta}
- 3\,\om_{\beta}\,H^{\alpha\beta} \nonumber \\
& & \hspace{25mm} - \ \eps^{\alpha\beta\gam}\,[\ \sig_{\beta\delta}
\,H^{\delta}\!_{\gam} - \sfrac{3}{2}\,\om_{\beta}
\,q_{\gam} + n_{\beta\delta}\,(E^{\delta}\!_{\gam}
+\sfrac{1}{2}\,\pi^{\delta}\!_{\gam})\ ]
\\ \nonumber \\
\l{onfdivh}
0 & = & (C_{5})^{\alpha} \ := \ (\p_{\beta} - 3\,a_{\beta})\,
(H^{\alpha\beta}) + (\mu+p)\,\om^{\alpha}
+ 3\,\om_{\beta}\,(E^{\alpha\beta}
-\sfrac{1}{6}\,\pi^{\alpha\beta}) - \sfrac{1}{2}\,
n^{\alpha}\!_{\beta}\,q^{\beta} \nonumber \\
& & \hspace{25mm} + \ \eps^{\alpha\beta\gam}\,[\ \sfrac{1}{2}
\,(\p_{\beta} - a_{\beta})\,(q_{\gam})
+ \sig_{\beta\delta}\,(E^{\delta}\!_{\gam}+\sfrac{1}{2}\,
\pi^{\delta}\!_{\gam}) - n_{\beta\delta}\,H^{\delta}\!_{\gam}\ ]
\\ \nonumber \\
\l{onfjac}
0 & = & (C_{J})^{\alpha} \ := \ (\p_{\beta} - 2\,a_{\beta})\,
(n^{\alpha\beta}) + \sfrac{2}{3}\,\Th\,\om^{\alpha}
+ 2\,\sig^{\alpha}\!_{\beta}\,\om^{\beta}
+ \eps^{\alpha\beta\gam}\,[\ \p_{\beta}(a_{\gam}) - 2\,\om_{\beta}
\,\Omega_{\gam}\ ]
\\ \nonumber \\
\l{onfgauss}
0 & = & (C_{G})^{\alpha\beta} \ := \ {}^{*}\!S^{\alpha\beta}
+ \sfrac{1}{3}\,\Th\,\sig^{\alpha\beta}
- \sig^{\lgl\alpha}\!_{\gam}\,\sigma^{\beta\rgl\gam}
- \om^{\lgl\alpha}\,\om^{\beta\rgl} + 2\,\om^{\lgl\alpha}\,
\Omega^{\beta\rgl} - (E^{\alpha\beta}+\sfrac{1}{2}\,
\pi^{\alpha\beta})
\\ \nonumber \\
\l{onffried}
0 & = & (C_{G}) \ := \ {}^{*}\!R + \sfrac{2}{3}\,\Th^{2}
- (\sig^{\alpha}\!_{\beta}\sig^{\beta}\!_{\alpha})
+ 2\,(\om_{\alpha}\om^{\alpha})
- 4\,(\om_{\alpha}\,\Omega^{\alpha}) - 2\,\mu - 2\,\Lambda \ ,
\eea
where
\bea
\l{onftf3ric}
{}^{*}\!S_{\alpha\beta} & := & \p_{\lgl\alpha}(a_{\beta\rgl})
+ b_{\lgl\alpha\beta\rgl} - \eps^{\gam\delta}{}_{\lgl\alpha}\,
(\p_{|\gam|} - 2\,a_{|\gam|})\,(n_{\beta\rgl\delta})
\\ \nonumber \\
\l{onf3rscl}
{}^{*}\!R & := &  2\,(2\,\p_{\alpha} - 3\,a_{\alpha})\,
(a^{\alpha}) - \sfrac{1}{2}\,b^{\alpha}\!_{\alpha}
\\ \nonumber \\
b_{\alpha\beta} & := & 2\,n_{\alpha\gam}\,n^{\gam}\!_{\beta}
- n^{\gam}\!_{\gam}\,n_{\alpha\beta} \ ,
\eea
and angle brackets denote the symmetric tracefree part. If
$\om^{\alpha} = 0$, such that the matter flow tangents ${\bf u}$
become the normals to a family of spacelike 3-surfaces ${\cal T}$:
$\left\{t=\mbox{const}\right\}$, the last two constraints in the
set correspond to the symmetric tracefree and trace parts of the
once-contracted Gau\ss\ embedding equation. In this case, one also
speaks of $(C_{G})$ as the generalised Friedmann equation, alias
the `Hamiltonian constraint' or the `energy constraint'. When
$\om^{\alpha} = 0$, the set of constraints should also include the
commutation relation Eq. (\r{onfcomss}).

\subsection{Evolution of spatial commutation functions}
The 9 spatial commutation functions $a^{\alpha}$ and
$n_{\alpha\beta}$ are generally evolved by Eqs. (40) and (41) given
in Ref. \ct{hveugg97}; these originate from the Jacobi
identity. Employing each of the constraints $(C_{1})^{\alpha}$ to
$(C_{3})^{\alpha\beta}$ listed in the previous paragraph, we can
eliminate $\p_{\alpha}$ frame derivatives of the kinematical fluid
variables $\Th$, $\sig_{\alpha\beta}$ and $\om^{\alpha}$ from their
right-hand sides. Thus, we obtain the following equations for the
evolution of the spatial commutation functions:
\bea
\l{onfadot}
\p_{0}(a^{\alpha}) & = & - \,\sfrac{1}{3}\,(\Th\,
\d^{\alpha}\!_{\beta} - \sfrac{3}{2}\,
\sig^{\alpha}\!_{\beta})\,(\udot^{\beta} + a^{\beta})
+ \sfrac{1}{2}\,n^{\alpha}\!_{\beta}\,\om^{\beta}
- \sfrac{1}{2}\,q^{\alpha} \nonumber \\
& & \hsp5 - \ \sfrac{1}{2}\,\eps^{\alpha\beta\gam}\,
[\ (\udot_{\beta} + a_{\beta}) - n_{\beta\delta}\,
\sig^{\delta}\!_{\gam} - (\p_{\beta}+\udot_{\beta}
-2\,a_{\beta})\,(\Omega_{\gam})\ ]
+ \sfrac{1}{2}\,(C_{1})^{\alpha}
\\ \nonumber \\
\l{onfndot}
\p_{0}(n^{\alpha\beta}) & = & - \,\sfrac{1}{3}\,\Th\,
n^{\alpha\beta} - \sig^{\lgl\alpha}\!_{\gam}\,
n^{\beta\rgl\gam} + \sfrac{1}{2}\,\sig^{\alpha\beta}\,
n^{\gam}\!_{\gam} - (\udot^{\lgl\alpha} + a^{\lgl\alpha})\,
\om^{\beta\rgl} - H^{\alpha\beta} + (\d^{\gam\lgl\alpha}\,
\p_{\gam} + \udot^{\lgl\alpha})\,(\Omega^{\beta\rgl})
\nonumber \\
& & \hsp5 - \ \sfrac{2}{3}\,\d^{\alpha\beta}\,
[\ 2\,(\udot_{\gam} + a_{\gam})\,\om^{\gam}
- \sig^{\gam}\!_{\delta}\,n^{\delta}\!_{\gam}
+ (\p_{\gam} + \udot_{\gam})\,(\Omega^{\gam})\ ]
\nonumber \\
& & \hsp5 - \ \eps^{\gam\delta\lgl\alpha}\,
[\ (\udot_{\gam} + a_{\gam})\,\sig^{\beta\rgl}\!_{\delta}
- (\om_{\gam}+ 2\,\Omega_{\gam})\,n^{\beta\rgl}\!_{\delta}\ ]
- \sfrac{2}{3}\,\d^{\alpha\beta}\,(C_{2}) + (C_{3})^{\alpha\beta}
\ .
\eea
%

\subsection{Evolution of kinematical fluid variables}
The evolution equations for the 9 kinematical fluid variables
$\Th$, $\sig_{\alpha\beta}$ and $\om^{\alpha}$ are provided by the
familiar Ricci field equations, i.e.
\bea
\l{onfthdot}
\p_{0}(\Th) - \p_{\alpha}(\udot^{\alpha})
& = & -\,\sfrac{1}{3}\,\Th^{2}
+ (\udot_{\alpha} - 2\,a_{\alpha})\,\udot^{\alpha}
- (\sig^{\alpha}\!_{\beta}\sig^{\beta}\!_{\alpha})
+ 2\,(\om_{\alpha}\om^{\alpha}) - \sfrac{1}{2}\,(\mu+3p) + \Lambda
\\ \nonumber \\
\l{onfsigdot}
\p_{0}(\sig^{\alpha\beta}) - \d^{\gam\lgl\alpha}\,
\p_{\gam}(\udot^{\beta\rgl})
& = & - \,\sfrac{2}{3}\,\Th\,\sig^{\alpha\beta}
+ (\udot^{\lgl\alpha} + a^{\lgl\alpha})\,\udot^{\beta\rgl}
- \sig^{\lgl\alpha}\!_{\gam}\,\sig^{\beta\rgl\gam}
- \om^{\lgl\alpha}\,\om^{\beta\rgl} 
- (E^{\alpha\beta}-\sfrac{1}{2}\,\pi^{\alpha\beta}) \nonumber \\
& & \hsp5 + \ \eps^{\gam\delta\lgl\alpha}\,[\ 2\,\Omega_{\gam}\,
\sig^{\beta\rgl}\!_{\delta} - n^{\beta\rgl}\!_{\gam}\,
\udot_{\delta}\ ]
\\ \nonumber \\
\l{onfomdot}
\p_{0}(\om^{\alpha}) - \sfrac{1}{2}\,\eps^{\alpha\beta\gam}\,
\p_{\beta}(\udot_{\gam})
& = & - \,\sfrac{2}{3}\,\Th\,\om^{\alpha} + \sig^{\alpha}\!_{\beta}
\,\om^{\beta} - \sfrac{1}{2}\,n^{\alpha}\!_{\beta}\,\udot^{\beta}
- \sfrac{1}{2}\,\eps^{\alpha\beta\gam}\,[\ a_{\beta}\,
\udot_{\gam} - 2\,\Omega_{\beta}\,\om_{\gam}\ ] \ .
\eea
%
\subsection{Evolution of Weyl curvature and matter variables}
Finally, we have the Bianchi field equations for the 10 Weyl
curvature variables $E_{\alpha\beta}$ and $H_{\alpha\beta}$ and the
4 matter variables $\mu$ and $q^{\alpha}$, which are obtained from
the once-contracted and twice-contracted second Bianchi identity,
respectively \ct{ell71,hve96,fri98}:
\bea
\l{onfedot}
\p_{0}(E^{\alpha\beta}+\sfrac{1}{2}\,\pi^{\alpha\beta})
- \eps^{\gam\delta\lgl\alpha}\,\p_{\gam}(H^{\beta\rgl}\!_{\delta})
& & \nonumber \\
\hsp5 + \ \sfrac{1}{2}\,\d^{\gam\lgl\alpha}\,\p_{\gamma}
(q^{\beta\rgl})
& = & - \,\sfrac{1}{2}\,(\mu+p)\,\sig^{\alpha\beta}
- \Th\,(E^{\alpha\beta}+\sfrac{1}{6}\,\pi^{\alpha\beta})
+ 3\,\sigma^{\lgl\alpha}\!_{\gam}\,(E^{\beta\rgl\gam}
-\sfrac{1}{6}\,\pi^{\beta\rgl\gam}) \nonumber \\
& & \hsp5 + \ \sfrac{1}{2}\,n^{\gam}\!_{\gam}\,H^{\alpha\beta}
- 3\,n^{\lgl\alpha}\!_{\gam}\,H^{\beta\rgl\gam}
- \sfrac{1}{2}\,(2\,\udot^{\lgl\alpha}
+ a^{\lgl\alpha})\,q^{\beta\rgl} \nonumber \\
& & \hsp5 + \ \eps^{\gam\delta\lgl\alpha}\,[\ (2\,\udot_{\gam}
- a_{\gam})\,H^{\beta\rgl}\!_{\delta} \nonumber \\
& & \hspace{20mm} + \ (\om_{\gam}+2\,\Omega_{\gam})\,
(E^{\beta\rgl}\!_{\delta}+\sfrac{1}{2}\,\pi^{\beta\rgl}\!_{\delta})
+ \sfrac{1}{2}\,n^{\beta\rgl}\!_{\gam}\,q_{\delta}\ ]
\\ \nonumber \\
\l{onfhdot}
\p_{0}(H^{\alpha\beta}) + \eps^{\gam\delta\lgl\alpha}\,
\p_{\gam}(E^{\beta\rgl}\!_{\delta}-\sfrac{1}{2}\,
\pi^{\beta\rgl}\!_{\delta})
& = & - \,\Th\,H^{\alpha\beta} + 3\,\sig^{\lgl\alpha}\!_{\gam}\,
H^{\beta\rgl\gam} + \sfrac{3}{2}\,\om^{\lgl\alpha}\,q^{\beta\rgl}
\nonumber \\
& & \hsp5 - \ \sfrac{1}{2}\,n^{\gam}\!_{\gam}\,(E^{\alpha\beta}
-\sfrac{1}{2}\,\pi^{\alpha\beta}) + 3\,n^{\lgl\alpha}\!_{\gam}\,
(E^{\beta\rgl\gam}-\sfrac{1}{2}\,\pi^{\beta\rgl\gam}) \nonumber \\
& & \hsp5 + \ \eps^{\gam\delta\lgl\alpha}\,[\ a_{\gam}\,
(E^{\beta\rgl}\!_{\delta}-\sfrac{1}{2}\,\pi^{\beta\rgl}\!_{\delta})
- 2\,\udot_{\gam}\,E^{\beta\rgl}\!_{\delta} \nonumber \\
& & \hspace{20mm} + \sfrac{1}{2}\,\sig^{\beta\rgl}\!_{\gam}
\,q_{\delta} + (\om_{\gam}+2\,\Omega_{\gam})\,
H^{\beta\rgl}\!_{\delta}\ ]
\\ \nonumber \\
\l{onfqdot}
\p_{0}(q^{\alpha}) + \d^{\alpha\beta}\,\p_{\beta}(p)
+ \p_{\beta}(\pi^{\alpha\beta})
& = & - \sfrac{4}{3}\,\Th\,q^{\alpha} - \sig^{\alpha}\!_{\beta}\,
q^{\beta} - (\mu+p)\,\udot^{\alpha} - (\udot_{\beta}
- 3\,a_{\beta})\,\pi^{\alpha\beta} \nonumber \\
& & \hsp5 - \ \eps^{\alpha\beta\gam}\,[\ (\om_{\beta}
-\Omega_{\beta})\,q_{\gam} - n_{\beta\delta}\,\pi^{\delta}\!_{\gam}
\ ]
\\ \nonumber \\
\l{onfmudot}
\p_{0}(\mu) + \p_{\alpha}(q^{\alpha}) & = & - \,\Th\,(\mu+p)
- 2\,(\udot_{\alpha} - a_{\alpha})\,q^{\alpha}
- (\sig^{\alpha}\!_{\beta}\pi^{\beta}\!_{\alpha}) \ .
\eea
\enl

The general extended $1+3$ ONF dynamical equations (\r{onfcomts})
and (\r{onfadot}) - (\r{onfmudot}) do not directly form a FOSH
evolution system, for two reasons. First, they do {\em not\/}
provide evolution equations for any of the geometrical variables
$\udot^{\alpha}$, $\Omega^{\alpha}$, $p$ and
$\pi_{\alpha\beta}$.\footnote{The time derivative of the latter is
indeed included in one of the Bianchi field equations, but not in a
`pure' form; because of the dynamical meaning of
$\pi_{\alpha\beta}$ discussed below, we regard this as an equation
for $E_{\alpha\beta}$.} In the case of $\Omega^{\alpha}$ this is a
reflection of the freedom of choice of a particular frame
$\{\,\p_{a}\,\}$. The indeterminacy of propagation of the remaining
three variables reveals the necessity of tying the description of
non-vacuum gravitational phenomena to a thermodynamical description
of the matter source fields. This may be achieved, for example, by
a phenomenological scheme for dissipative relativistic fluids or a
relativistic kinetic theory approach. Second, the combinations of
derivatives occuring in the above equations do not have the
required symmetric structure. Given our present goal, the target is
to choose a suitable matter description and then shuffle the
evolution equations for the full set of variables into a FOSH form.

\subsection{New set of frame variables for rank $2$ symmetric
tracefree tensors}
In practical applications of the $1+3$ ONF formalism it proves
helpful to introduce a new set of variables adapted to the
tracefree condition $A^{\alpha}\!_{\alpha} = 0$ and the magnitude
$A^{2} := \sfrac{1}{2}\, A^{\alpha}\!_{\beta}\,A^{\beta}\!_{\alpha}
\geq 0$ of any $1+3$ invariantly defined rank 2 symmetric tracefree
tensor field ${\bf A}$ orthogonal to ${\bf u}$. In our case we
have, in particular, $A_{\alpha\beta} \in \{\,\sig_{\alpha\beta},
\,E_{\alpha\beta}, \,H_{\alpha\beta},
\,\pi_{\alpha\beta}\,\}$. This new set of irreducible frame
components is defined by (\,cf. Ref. \ct{waihsu89,hveugg97}\,)
\bea
\l{apm}
\begin{array}{lllll}
A_{+} :=  - \,{\sfrac{3}{2}}\,A_{11} 
= {\sfrac{3}{2}}\,(A_{22}+A_{33}) & &
A_{-} := {\sfrac{\sqrt{3}}{2}}\,(A_{22}-A_{33}) & & \\ \\
A_{1} := \sqrt{3}\,A_{23} & &
A_{2} := \sqrt{3}\,A_{31} & &
A_{3} := \sqrt{3}\,A_{12} \ .
\end{array}
\eea
It has the property that the magnitude of ${\bf A}$ assumes the
explicit form
\be
A^{2} = \sfrac{1}{3}\,[\ (A_{+})^{2} + (A_{-})^{2} + (A_{1})^{2}
+ (A_{2})^{2} + (A_{3})^{2}\ ] \ .
\ee
It should be emphasised that this decomposition arbitrarily adapts
to the spatial ${\bf e}_{1}$-axis. However, this is only a matter
of convention, and by a cyclic permutation of indices $1
\rightarrow 2 \rightarrow 3 \rightarrow 1$ one can easily adapt to
any of the other spatial axes.

In principle, one can extend this procedure to also include the
spatial commutation functions $n_{\alpha\beta}$, i.e., splitting
them into a trace and a tracefree part, where the latter is further
subdivided according to the scheme above. However, as
$n_{\alpha\beta}$ has no immediate invariant meaning in the $1+3$
covariant picture, this possibility will not be pursued any further
here.

\section{$\left(\,{\cal M},\,{\bf g},\,{\bf u}\,\right)$ with
barotropic perfect fluid matter source fields}
\l{sec:foshpf}
We now specialise the energy-momentum-stress tensor of the matter
source fields appearing in the dynamical equations of section
\r{sec:13eqs} to a perfect fluid form (with matter flow tangents
${\bf u}$) by setting
\be
\l{pf}
0 = q^{\alpha} = \pi_{\alpha\beta}
\ee
throughout,\footnote{We also discard the cosmological constant,
$\Lambda = 0$.} and we assume a barotropic equation of state, i.e.,
a functional dependence of the isotropic fluid pressure on the
energy density only,
\be
\l{eos}
p = p(\mu) \ .
\ee
We define $c_{s}^{2}(\mu) := dp(\mu)/d\mu$ as the isentropic {\em
speed of sound\/}. Under the assumptions (\r{pf}), the evolution
equation for $q^{\alpha}$, Eq. (\r{onfqdot}), reduces to the {\em
new\/} constraint
\be
\l{mom}
0 = (C_{PF})^{\alpha} := \d^{\alpha\beta}\,\p_{\beta}(p)
+ (\mu+p)\,\udot^{\alpha} \ ,
\ee
which is often called the momentum conservation equation.

\subsection{Derivation of a FOSH evolution system}
By applying the commutator (\r{onfcomts}) to $f = p$ and using
Eqs. (\r{eos}), (\r{mom}), (\r{onfmudot}) and (\r{onfdivsig}), we
derive an evolution equation for the acceleration $\udot^{\alpha}$
of the matter fluid tangents ${\bf u}$ given by
\bea
\l{onfpfudot}
& & \p_{0}(\udot^{\alpha}) - c_{s}^{2}\,\p_{\beta}(\sfrac{1}{3}\,
\Th\,\d^{\alpha\beta} + \sig^{\alpha\beta}
+ \eps^{\alpha\beta\gam}\,\om_{\gam}) \nonumber \\
& & \hspace{35mm} = -\,[\ c_{s}^{-2}\,\frac{d^{2}p}{d\mu^{2}}\,
(\mu+p) - c_{s}^{2} + \sfrac{1}{3}\ ]\ \Th\,\udot^{\alpha}
- (\udot_{\beta}+3c_{s}^{2}a_{\beta})\,\sig^{\alpha\beta}
- c_{s}^{2}\,n^{\alpha}\!_{\beta}\,\om^{\beta}
\nonumber \\
& & \hspace{40mm} + \ \eps^{\alpha\beta\gam}\,[\ (2c_{s}^{2}-1)\,
\om_{\beta}\,\udot_{\gam} + \Omega_{\beta}\,\udot_{\gam}
- c_{s}^{2}\,a_{\beta}\,\om_{\gam} - c_{s}^{2}\,n_{\beta\delta}\,
\sig^{\delta}\!_{\gam}\ ]
\nonumber \\
& & \hspace{40mm} - \ c_{s}^{2}\,(C_{1})^{\alpha} + \Th\,
c_{s}^{-2}\,\frac{d^{2}p}{d\mu^{2}}\,(C_{PF})^{\alpha} \ .
\eea
Here we have suppressed an evolution equation for
$(C_{PF})^{\alpha}$ which arises along the way. Next, contracting
the commutator (\r{onfcomss}), again applied to $f = p$, with
$\eps^{\alpha\beta\gam}$ and using Eqs. (\r{eos}), (\r{onfmudot})
and (\r{mom}) leads to the identity\footnote{In $1+3$ covariant
terms this is just $0 = \eps^{abc}\,\D_{b}\udot_{c}
-2\,c_{s}^{2}\,\Th\,\om^{a} - \dots$ (\,see Ref. \ct{hve96},
Ch.2\,).}
\bea
\l{curlacc}
0 & = & \eps^{\alpha\beta\gam}\,(\p_{\beta}-a_{\beta})\,
(\udot_{\gam}) - n^{\alpha}\!_{\beta}\,\udot^{\beta} -
2\,c_{s}^{2}\,\Th\,\om^{\alpha} \nonumber \\
\nonumber \\
& & \hsp5 - \ (\mu+p)^{-1}\,[\ \eps^{\alpha\beta\gam}\,
(\p_{\beta}-a_{\beta})\,(C_{PF})_{\gam} -
n^{\alpha}\!_{\beta}\,(C_{PF})^{\beta} + (c_{s}^{-2}
+1)\,\eps^{\alpha\beta\gam}\,\udot_{\beta}\,(C_{PF})_{\gam} \ ]
\ . 
\eea
This identity constitutes the {\em key step\/} in achieving FOSH
form for the evolution subsystem (\r{onfthdot}) - (\r{onfomdot})
and (\r{onfpfudot}) that links the kinematical fluid variables
$\Th$, $\sig_{\alpha\beta}$ and $\om^{\alpha}$ to $\udot^{\alpha}$
and establishes the sound cone structure on $\left(\,{\cal
M},\,{\bf g},\,{\bf u}\,\right)$. The trick is to add, on using
Eq. (\r{curlacc}),
$\eps^{\alpha\beta\gam}\,\p_{\beta}(\udot_{\gam})$ to the left-hand
side of Eq. (\r{onfomdot}), i.e., to change its principle part to
the new form $\p_{0}(\om^{\alpha}) +
\sfrac{1}{2}\,\eps^{\alpha\beta\gam}\, \p_{\beta}(\udot_{\gam})$.
\enl

Having adopted the Lagrangean viewpoint for the description of the
dynamics by making the identification $\p_{0} = {\bf u}$, we now
proceed to introduce a set of (dimensionless) comoving local
coordinates $\{\,x^{\mu}\,\} = \{\,t, \,x^{i}\,\}$. The natural way
of doing this is via the $1+3$ {\em threading approach\/} discussed
in detail by Jantzen et al \ct{janetal92} and Boersma and Dray
\ct{boedra95}. Defining coordinate components of the $\p_{a}$ by
$e_{a}{}^{\mu} := \p_{a}(x^{\mu})$, the frame spanning basis fields
can be expressed as \ct{hveugg97}
\be
\l{onf13}
\p_{0} = e_{0}{}^{\mu}\,\ptl_{\mu}
:= M^{-1}\,\ptl_{t} \ , \hsp5
\p_{\alpha} = e_{\alpha}{}^{\mu}\,\ptl_{\mu}
:= e_{\alpha}{}^{i}\,(M_{i}\,\ptl_{t}+\ptl_{i}) \ ,
\ee
where $M = M(t,x^{i})$ is the {\em threading lapse function\/} and
$M_{i}\,dx^{i} = M_{i}(t,x^{j})\,dx^{i}$ is the dimensionless {\em
threading shift 1-form\/}. The inverse of the {\em threading
metric\/} is $h^{ij} :=
\d^{\alpha\beta}\,e_{\alpha}{}^{i}\,e_{\beta}{}^{j}$. Note the
presence of the temporal partial derivative in the `spatial' frame
fields $\p_{\alpha}$. With the expanded frame fields (\r{onf13})
inserted, the commutator equations (\r{onfcomts}) and
(\r{onfcomss}) now yield \ct{hveugg97}
\bea
\l{onftacc}
e_{\alpha}{}^{i}\,[\ \ptl_{t}M_{i} + M^{-1}\,(M_{i}\,\ptl_{t}M
+ \ptl_{i}M)\ ] & = & \udot_{\alpha} \\ \nonumber \\
\l{onfttriadevol}
M^{-1}\,\ptl_{t}e_{\alpha}{}^{i} & = & - \,[\ \sfrac{1}{3}\,\Th\,
\d_{\alpha}{}^{\beta} + \sig_{\alpha}{}^{\beta}
+ \eps_{\alpha}{}^{\beta}{}_{\gam}\,
(\om^{\gamma}+\Omega^{\gamma})\ ]\ e_{\beta}{}^{i} \\ \nonumber \\
\l{onftvor}
M\,e_{[\alpha}{}^{i}\,e_{\beta]}{}^{j}\,(M_{i}\,\ptl_{t}M_{j}
+ \ptl_{i}M_{j})
& = & \eps_{\alpha\beta\gamma}\,\om^{\gamma}
\\ \nonumber \\
\l{onfttconstr}
2\,e_{[\alpha}{}^{i}\,[\ M_{i}\,\ptl_{t}e_{\beta]}{}^{j}
+ \ptl_{i}e_{\beta]}{}^{j}\ ]\,e^{\gamma}{}_{j}
& = & 2\,a_{[\alpha}\,\delta^{\gamma}\!_{\beta]}
+ \epsilon_{\alpha\beta\delta}\,n^{\delta\gamma} \ ,
\eea
where $e^{\alpha}{}_{i}$ is defined through the relation
$e^{\alpha}{}_{i}\,e_{\alpha}{}^{j} = \delta^{j}{}_{i}$. The
threading shift 1-form will be {\em non-zero\/}, $M_{i} \neq 0$, if
the worldlines of the matter fluid are {\em rotating\/}, i.e.,
$\om^{\alpha} \neq 0$. In this case the local rest 3-spaces do {\em
not\/} mesh together to form a family of spacelike 3-surfaces
${\cal T}$: $\left\{t=\mbox{const}\right\}$ everywhere orthogonal
to ${\bf u}$. Substitution of $\ptl_{t}M_{j}$ in Eq. (\r{onftvor})
from Eq. (\r{onftacc}) generates a further constraint relation to
be satisfied. \enl

One way of picking the threading lapse function $M$ is provided by
parameterising the integral curves of ${\bf u}$ by physical {\em
proper time\/}. Then we have
\be
\l{proper}
e_{0}{}^{\mu} = u^{\mu} = M_{0}^{-1}\,\d^{\mu}\!_{0} \ ,
\ee
where $M_{0}$ denotes a constant threading lapse function of {\em
unit\/} length. If, furthermore, the spatial frame
$\{\,\p_{\alpha}\,\}$ is chosen to be {\em Fermi-propagated\/}
along ${\bf u}$, i.e.,
\be
\l{fermi}
\Omega^{\alpha} = 0 \ ,
\ee
we obtain from Eqs. (\r{onftacc}) and (\r{onfttriadevol})
reduced evolution equations for $M_{i}$ and $e_{\alpha}{}^{i}$;
\bea
\l{tsevo}
\ptl_{t}M_{i} & = & \udot_{\alpha}\,e^{\alpha}{}_{i} = \udot_{i}
\\ \nonumber \\
\l{fccevo}
M_{0}^{-1}\,\ptl_{t}e_{\alpha}{}^{i} & = & - \,[\ \sfrac{1}{3}\,
\Th\,\d_{\alpha}{}^{\beta} + \sig_{\alpha}{}^{\beta}
+ \eps_{\alpha}{}^{\beta}{}_{\gam}\,\om^{\gamma}\ ]
\ e_{\beta}{}^{i} \ .
\eea

An alternative to a constant threading lapse function and a
dynamical threading shift 1-form is the prescription for barotropic
perfect fluids given by Salzman and Taub \ct{saltau54}, where
\be
\l{threadm}
M = M_{0}\,\exp\left[\ -\int_{p_{0}}^{p}\,\frac{dp'}{(\mu+p')}
\ \right] \ , \hsp5 M_{i} = M_{i}(x^{j}) \ ;
\ee
here $M$ is dynamical while $M_{i}$ is coordinate time
independent. This choice reduces the number of dependent field
variables by three. By Eq. (\r{onftvor}) the time dependence of
$\om^{\alpha}$ then lives exclusively in $e_{\alpha}{}^{i}$, which
is determined by Eq. (\r{onfttriadevol}). \enl

Our preparations for the derivation of an evolution system in FOSH
form for barotropic perfect fluids from the general $1+3$ ONF
equations are now complete. To this end we will use the set given
by the matter content specifying equations (\r{pf}) and (\r{eos}),
the frame fixing equations (\r{proper}) and (\r{fermi}), the
evolution equations (\r{tsevo}), (\r{fccevo}), (\r{onfadot}) -
(\r{onfhdot}), (\r{onfmudot}) and (\r{onfpfudot}), and the identity
(\r{curlacc}). Of immediate interest for the FOSH structure is only
the principle part of the evolution equations, the left-hand side
in Eq. (\r{fosh}), that describes the dynamical interactions
between the various fields. In terms of the frame derivatives
$\p_{a}$ we can represent it by \enl

\noindent
{\em The frame derivative principle part\/}:
\be
\l{pp}
\bar{M}{}^{AB\,a}\,\p_{a}(u_{B}) \ .
\ee
Using tracefree-adapted irreducible frame variables as defined in
Eq. (\r{apm}) for each of the fluid rate of shear and the electric
and magnetic Weyl curvature, a FOSH evolution system can now be
obtained, taking certain linear combinations of the equations where
necessary, for the following set of 44 dependent dynamical fields:
\enl

\noindent
{\em The dependent geometrical field variables\/}: \nopagebreak
\bea
\l{depvar}
u^{A} =
\left( \begin{array}{l}
u_{frame} \\ u_{3\mbox{-}con} \\ u_{kin,1} \\  u_{kin,2} \\
u_{kin,3} \\ u_{mat} \\ u_{Weyl}
        \end{array} \right) \ , \hsp5
\begin{array}{lcl}
u_{frame} & = & [\ e_{\alpha}{}^{i}, \,M_{i}\ ]^{T} \\
u_{3\mbox{-}con} & = & [\ a^{\alpha}, \,n_{\alpha\beta}\ ]^{T}
\\
u_{kin,1} & = & [\ \udot_{1}, \,\sfrac{1}{3}(\Th-2\sig_{+}),
\,(\sfrac{1}{\sqrt{3}}\sig_{3}+\om_{3}),
\,(\sfrac{1}{\sqrt{3}}\sig_{2}-\om_{2})\ ]^{T} \\
u_{kin,2} & = & [\ \udot_{2}, \,\sfrac{1}{3}(\Th+\sig_{+}
+\sqrt{3}\sig_{-}), \,(\sfrac{1}{\sqrt{3}}\sig_{1}+\om_{1}),
\,(\sfrac{1}{\sqrt{3}}\sig_{3}-\om_{3})\ ]^{T} \\
u_{kin,3} & = & [\ \udot_{3}, \,\sfrac{1}{3}(\Th+\sig_{+}
-\sqrt{3}\sig_{-}), \,(\sfrac{1}{\sqrt{3}}\sig_{2}+\om_{2}),
\,(\sfrac{1}{\sqrt{3}}\sig_{1}-\om_{1})\ ]^{T} \\
u_{mat} & = & [\ \mu\ ] \\
u_{Weyl} & = & [\ E_{+}, \,E_{-}, \,E_{1}, \,E_{2}, \,E_{3},
\,H_{+}, \,H_{-}, \,H_{1}, \,H_{2}, \,H_{3}\ ]^{T}
\end{array} \ .
\eea
The symmetric $(44\times 44)$-matrices $\bar{M}{}^{AB\,a}$
occurring in Eq. (\r{pp}) are found to assume the forms \enl

\noindent
{\em The matrices\/}:
\bea
\l{mab01}
\bar{M}{}^{AB\,0} =
\left( \begin{array}{ccccccc}
        {\bf 1}_{12} & . & . & . & . & . & . \\
        . & {\bf 1}_{9} & . & . & . & . & . \\
        . & . & {\bf K} & . & . & . & . \\
        . & . & . & {\bf K} & . & . & . \\
        . & . & . & . & {\bf K} & . & . \\
        . & . & . & . & . & 1 & . \\
        . & . & . & . & . & . & {\bf 1}_{10}
       \end{array}
\right) \ , \hsp5
\bar{M}{}^{AB\,1} =
\left( \begin{array}{ccccccc}
        {\bf 0}_{12} & . & . & . & . & . & . \\
        . & {\bf 0}_{9} & . & . & . & . & . \\
        . & . & {\bf B}_{1} & . & . & . & . \\
        . & . & . & {\bf B}_{2} & . & . & . \\
        . & . & . & . & {\bf B}_{3} & . & . \\
        . & . & . & . & . & 0 & . \\
        . & . & . & . & . & . & {\bf C}_{1}
       \end{array}
\right) \ ,
\eea
\bea
\l{mab23}
\bar{M}{}^{AB\,2} =
\left( \begin{array}{ccccccc}
        {\bf 0}_{12} & . & . & . & . & . & . \\
        . & {\bf 0}_{9} & . & . & . & . & . \\
        . & . & {\bf B}_{3} & . & . & . & . \\
        . & . & . & {\bf B}_{1} & . & . & . \\
        . & . & . & . & {\bf B}_{2} & . & , \\
        . & . & . & . & . & 0 & . \\
        . & . & . & . & . & . & {\bf C}_{2}
       \end{array}
\right) \ , \hsp5
\bar{M}{}^{AB\,3} =
\left( \begin{array}{ccccccc}
        {\bf 0}_{12} & . & . & . & . & . & . \\
        . & {\bf 0}_{9} & . & . & . & . & . \\
        . & . & {\bf B}_{2} & . & . & . & . \\
        . & . & . & {\bf B}_{3} & . & . & . \\
        . & . & . & . & {\bf B}_{1} & . & . \\
        . & . & . & . & . & 0 & . \\
        . & . & . & . & . & . & {\bf C}_{3}
       \end{array}
\right) \ .
\eea
\bea
\l{kinsec}
{\bf K} :=
\left( \begin{array}{cc}
        1 & .                     \\
        . & c_{s}^{2}\,{\bf 1}_{3}
       \end{array}
\right) \ , \hsp5
{\bf B}_{1} := -\,c_{s}^{2}
\left( \begin{array}{cccc}
        . & 1 & . & . \\
        1 & . & . & . \\
        . & . & . & . \\
        . & . & . & .
       \end{array}
\right) \ , \hsp5
{\bf B}_{2} := -\,c_{s}^{2}
\left( \begin{array}{cccc}
        . & . & . & 1 \\
        . & . & . & . \\
        . & . & . & . \\
        1 & . & . & .
       \end{array}
\right) \ ,
{\bf B}_{3} := -\,c_{s}^{2}
\left( \begin{array}{cccc}
        . & . & 1 & . \\
        . & . & . & . \\
        1 & . & . & . \\
        . & . & . & .
       \end{array}
\right) \ .
\eea
\bea
{\bf C}_{1} :=
\left( \begin{array}{cccccccccc}
        . & . & . & . & . & . & .    & . & .   & .      \\
        . & . & . & . & . & . & .    & 1 & .   & .      \\
        . & . & . & . & . & . & -\,1 & . & .   & .      \\
        . & . & . & . & . & . & .    & . & .   & -\,1/2 \\
        . & . & . & . & . & . & .    & . & 1/2 & .      \\
        . & . & .    & .      & .   & . & . & . & . & . \\
        . & . & -\,1 & .      & .   & . & . & . & . & . \\
        . & 1 & .    & .      & .   & . & . & . & . & . \\
        . & . & .    & .      & 1/2 & . & . & . & . & . \\
        . & . & .    & -\,1/2 & . & . & . & . & . & . \\
       \end{array}
\right) \ ,
\eea
\bea
{\bf C}_{2} :=
\left( \begin{array}{cccccccccc}
        . & . & . & . & . & . & . & . & \sqrt{3}/2 & .   \\
        . & . & . & . & . & . & . & . & -\,1/2     & .   \\
        . & . & . & . & . & . & . & . & . & 1/2 \\
        . & . & . & . & . & -\,\sqrt{3}/2 & 1/2 & .      & . & . \\
        . & . & . & . & . & .             & .   & -\,1/2 & . & . \\
        . & . & . & -\,\sqrt{3}/2 & . & . & . & . & . & . \\
        . & . & . & 1/2           & . & . & . & . & . & . \\
        . & . & . & . & -\,1/2 & . & . & . & . & . \\
        \sqrt{3}/2 & -\,1/2 & .   & . & . & . & . & . & . & . \\
        .          & .      & 1/2 & . & . & . & . & . & . & . \\
       \end{array}
\right) \ ,
\eea
\bea
{\bf C}_{3} :=
\left( \begin{array}{cccccccccc}
        . & . & . & . & . & . & . & . & . & -\,\sqrt{3}/2 \\
        . & . & . & . & . & . & . & . & . & -\,1/2        \\
        . & . & . & . & . & . & . & . & -\,1/2 & . \\
        . & . & . & . & . & .          & .   & 1/2 & . & . \\
        . & . & . & . & . & \sqrt{3}/2 & 1/2 & .   & . & . \\
        . & . & . & . & \sqrt{3}/2 & . & . & . & . & . \\
        . & . & . & . & 1/2        & . & . & . & . & . \\
        . & . & . & 1/2 & . & . & . & . & . & . \\
        .             & .      & -\,1/2 & . & . & . & . & . & . & . \\
        -\,\sqrt{3}/2 & -\,1/2 & .      & . & . & . & . & . & . & . \\
       \end{array}
\right) \ .
\eea
With the exception of the submatrices ${\bf K}$ and ${\bf B}_{1}$
to ${\bf B}_{3}$ in the kinematical branch of our FOSH evolution
system that contain $c_{s}^{2} = c_{s}^{2}(\mu)$, the remaining
components of the $\bar{M}{}^{AB\,a}$ are just numerical constants,
since we evaluated the principle part in the $1+3$ ONF basis
$\{\,\p_{a}\,\}$ (and because of reasons which we will briefly come
back to below). When the frame derivatives $\p_{a}$ in Eq. (\r{pp})
are substituted for in terms of their coordinate components and the
partial derivatives introduced in Eqs. (\r{onf13}) and
(\r{proper}), the non-constant components of the symmetric matrices
$M^{AB\,\mu}$ in Eq. (\r{fosh}) with respect to the local
coordinate basis $\{\,\ptl_{\mu}\,\}$ can be easily read
off. Finally, hyperbolicity will follow from the reality of the
eigenvalues we obtain when determining the characteristic
3-surfaces in the next subsection.

\subsection{Characteristic 3-surfaces, propagation velocities and
eigenfields}
The set of characteristic 3-surfaces $\{\phi=\mbox{const}\}$
underlying a FOSH evolution system can be interpreted as a
collection of wavefronts with phase function $\phi$ across which
certain physical quantities may be discontinuous. The associated
characteristic eigenfields propagate along so-called
bicharacteristic rays with tangents ${\bf k}$ {\em within\/} these
3-surfaces at velocities $v$, where $v = \tanh\alpha$ is a function
of the hyperbolic angle of tilt $\alpha$ between ${\bf k}$ and
${\bf u}$ \ct{couhil62}. To determine the characteristic
3-surfaces, we can assume without loss of generality that at a
particular spacetime event the $1+3$ ONF is oriented in such a way
that the orthogonal gradient vector fields $\mbox{\boldmath
$\xi$}$, $\xi_{a} := \nabla_{a}\phi$, are tangent to a 2-surfaces
spanned by ${\bf u}$ and $\p_{1}$. This choice of frame, which then
only leaves the freedom of spatial rotations about the
$\p_{1}$-direction, can be made due to the local isotropy of the
characteristic cones (\,see, e.g.,
Refs. \ct{abretal97,fri96,hislin83}\,). Hence, locally the frame
axes and the given coordinate lines are made to coincide. In
particular, at a point we can take coordinate components of
$\p_{1}$ where $0 = e_{1}{}^{2} = e_{1}{}^{3}$ and also $M_{i} =
0$;\footnote{The vorticity of the matter flow lines cannot be felt
at a single point.} however, generically their time derivatives
will be non-zero. Thus, we obtain
\be
\l{onf132d}
\p_{0} = M_{0}^{-1}\,\ptl_{t} \ , \hsp5
\hsp5 \p_{1} = e_{1}{}^{1}\,\ptl_{1} \ .
\ee
%

\subsubsection{The characteristic condition}
The characteristic condition the vector fields $\mbox{\boldmath
$\xi$}$ have to satisfy is
\be
0 = \det\,[\,M^{AB\,\mu}\,\xi_{\mu}\,] \ .
\ee
With the choice of frame outlined, their coordinate components are
given by $\xi_{\mu} = -\,v\,u_{\mu} + e^{1}{}_{\mu}$
\ct{couhil62,gerlin91,hislin83}, where the parameter $v$ coincides
with the different possible propagation velocities of the
characteristic eigenfields\footnote{It also gives the slope of
$\mbox{\boldmath $\xi$}$ with respect to $\p_{1}$.} and
$e^{1}{}_{\mu}$ denotes the inverse coordinate components of
$\p_{1}$. With Eq. (\r{onf132d}) we find, therefore, $\xi_{\mu} =
v\,M_{0}\,\d^{0}\!_{\mu} + (e_{1}{}^{1})^{-1}\,\d^{1}\!_{\mu}$,
leading to \enl

\bea
0 & = & \det\,[\,v\,M_{0}\,M^{AB\,0}
+ (e_{1}{}^{1})^{-1}\,M^{AB\,1}\,] \nonumber
\\ \nonumber \\
& = & (M_{0})^{44}\,c_{s}^{6}\,v^{30}\,
(v-c_{s})^{3}\,(v+c_{s})^{3}\,
(v-1)^{2}\,(v+1)^{2}\,(v-\sfrac{1}{2})^{2}\,(v+\sfrac{1}{2})^{2}
\ . 
\eea
Clearly {\em all\/} roots $v$ of this equation are real
(\,establishing hyperbolicity; cf. Ref. \ct{couhil62}\,), and {\em
no\/} root $v$ is single-valued, implying a set of characteristic
3-surfaces each of which is degenerate.

\subsubsection{Characteristic velocities and eigenfields}
First, we find 30 characteristic eigenfields of our
FOSH evolution system that propagate with velocity $v_{1} = 0$ with
respect to observers comoving with ${\bf u}$; these are
\be
\l{chef1}
u_{(1)}^{A} = [\ e_{\alpha}{}^{i}, \,M_{i}, \,a^{\alpha},
\,n_{\alpha\beta}, \,\sfrac{1}{3}(\Th+\sig_{+}\pm\sqrt{3}\sig_{-}),
\,(\sfrac{1}{\sqrt{3}}\sig_{1}\pm\om_{1}),
\,(\sfrac{1}{\sqrt{3}}\sig_{2}-\om_{2}),
\,(\sfrac{1}{\sqrt{3}}\sig_{3}+\om_{3}), \,\mu, \,E_{+},
\,H_{+}\ ]^{T} \ .
\ee

Second, upon diagonalisation of the $\p_{1}$-adapted, reduced
principle part in Eq. (\r{fosh}), the set of 6 characteristic
eigenfields propagating with velocities $v_{2,3} = \pm\,c_{s}$
along the sound cone is found to be
\be
\l{chef2}
u_{(2)}^{A} = [\ \sfrac{1}{\sqrt{1+c_{s}^{2}}}\,(\,c_{s}\udot_{1}
\pm\sfrac{1}{3}(\Th-2\sig_{+})\,), \,\sfrac{1}{\sqrt{1+c_{s}^{2}}}
\,(\,c_{s}\udot_{2}\pm(\sfrac{1}{\sqrt{3}}\sig_{3}-\om_{3})\,),
\,\sfrac{1}{\sqrt{1+c_{s}^{2}}}\,(\,c_{s}\udot_{3}\pm
(\sfrac{1}{\sqrt{3}}\sig_{2}+\om_{2})\,)\ ]^{T} \ .
\ee
Here, and in the following, the upper sign applies to outgoing
modes and the lower one to incoming modes.\footnote{In
observational cosmology, where we are faced with a situation of
`here and now' in spacetime, it is the incoming modes that are of
physical relevance.} By construction, the tracefree-adapted
variables of Eq. (\r{apm}) clearly exhibit the purely {\em
longitudinal\/} character of the first two eigenfields\footnote{The
quantity $\sfrac{1}{3}(\Th-2\sig_{+})$ is just the longitudinal
component $\sfrac{1}{3}\,\Th + \sig_{11}$ of the rate of expansion
tensor.} and the semi-longitudinal of the latter four with respect
to the (assumed) spatial propagation direction $\p_{1}$. In the
pressure-free case, $p = 0 \Rightarrow \udot^{\alpha} = 0$, $c_{s}
= 0$, the remaining kinematical fluid variables $\Theta$,
$\sig_{\alpha\beta}$ and $\om^{\alpha}$ also propagate with $v_{1}
= 0$.\footnote{When $\udot^{\alpha} = 0$, Eqs. (\r{onfthdot}) -
(\r{onfsigdot}) do {\em not\/} need to be multiplied by $c_{s}^{2}$
to obtain a symmetric structure for the principle part in the
kinematical branch.}

Third, the set of 4 characteristic eigenfields propagating with velocities
$v_{4,5} = \pm\,1$ along the null cone is
\be
\l{chef3}
u_{(3)}^{A} = [\ \sfrac{1}{\sqrt{2}}\,(\,E_{-}\mp H_{1}\,), 
\,\sfrac{1}{\sqrt{2}}\,(\,E_{1}\pm H_{-}\,)\ ]^{T} \ ;
\ee
each pair corresponds to one of the two possible polarisation
states of the freely propagating gravitational field. Again, the
tracefree-adapted variables of Eq. (\r{apm}) nicely reveal that
these eigenfields are purely {\em transverse\/} to the (assumed)
spatial propagation direction $\p_{1}$. The full propagation
equations of these modes are given in appendix \r{app2}; in
appendix \r{app4} one can find expansions of all Weyl curvature
characteristic eigenfields using the gauge-invariant variables of
Bardeen for linearised perturbations of FLRW cosmological models
\ct{bar80}.

Finally one obtains 4 characteristic eigenfields propagating with
velocities $v_{6,7} = \pm\,\sfrac{1}{2}$ along timelike 3-surfaces
which are
\be
\l{chef4}
u_{(4)}^{A} = [\ \sfrac{1}{\sqrt{2}}\,(\,E_{3}\mp H_{2}\,),
\,\sfrac{1}{\sqrt{2}}\,(\,E_{2}\pm H_{3}\,)\ ]^{T} \ .
\ee
The tracefree-adapted variables of Eq. (\r{apm}) classify them as
{\em longitudinal\/} modes of the Weyl curvature, which also come
in two possible polarisation states. They have been identified
before by, e.g., Szekeres \ct{sze65,sze66}. The full propagation
equations of these modes are given in appendix \r{app2}. Their
physical relevance will be further discussed below. \enl

The results of Eqs. (\r{chef1}) - (\r{chef4}) lead to a number of
observations. The spacetime metric ${\bf g}$, which embodies the
local causal structure of $\left(\,{\cal M},\,{\bf g},\,{\bf
u}\,\right)$ (and has coordinate components constructed by
$g_{\mu\nu} = \eta_{ab}\,e^{a}{}_{\mu}\, e^{b}{}_{\nu}$),
propagates along the timelike reference congruence ${\bf u}$
itself, i.e., with $v = 0$ (\,see also Ref. \ct{putear96}\,). Parts
of the spacetime connection (which contains first derivatives of
${\bf g}$) also propagate at $v = 0$, while the remaining parts
follow the sound cone. With the present geometrical set of
dependent field variables, it is only within the Weyl curvature
(which is of second order in the derivatives of ${\bf g}$) that one
finds modes that propagate {\em changes\/} in the state of a
gravitational field at the speed of light.

As to the magnitudes of the different propagation velocities it
should be noted that, by use of the Bianchi field equations
(\r{onfedot}) and (\r{onfhdot}), {\em causal\/} propagation of the
gravitational field modes with $|v| \in \{\,0, \,\sfrac{1}{2},
\,1\,\}$ falls out automatically; no further assumptions are
required. To ensure {\em causal\/} propagation of pressure
perturbations in the matter fluid, on the other hand, we need to
impose the condition $0 \leq c_{s} < 1$.

For vacuum gravitational fields as well as perfect fluid spacetime
geometries (in matter comoving description), the different
characteristic 3-surfaces associated with the Weyl curvature do
{\em not\/} intersect due to the `linear degeneracy' of the
principle part in the Bianchi field equations. This means it is
{\em not genuinely\/} non-linear and, therefore, discontinuities
(`shocks') cannot form in the process of evolution. The only Weyl
curvature discontinuities that propagate are those contained in the
initial data. The situation is different in the sound cone branch
of our FOSH evolution system, where the (non-constant, equation of
state dependent) characteristic velocity $c_{s}$ explicitly occurs
in the related eigenfields given in Eq. (\r{chef2}), and shock
formation is possible (\,on these issues see, e.g., Bona in
Ref. \ct{bon96}\,). A different phenomenon altogether are the
`shell-crossing singularities'. At these events the Lagrangean
description breaks down because integral curves of ${\bf u}$
intersect.

\subsubsection{Polarisation states and the NP formalism}
We mentioned before that by our method of choosing a $1+3$ ONF
$\{\,\p_{a}\,\}$, with $\p_{0}$ aligned along the uniquely defined
matter flow lines and $\p_{1}$ locally aligned along an arbitrarily
chosen propagation direction (for the purpose of determining the
characteristic structure of our FOSH evolution system), the
remaining frame freedom, given this choice of propagation
direction, consists of spatial rotations about the
$\p_{1}$-direction through an angle $\varphi$ only. Focusing on the
characteristic eigenfields of the Weyl curvature, this implies that
out of, say, the 4 outgoing modes, in general we can set only one
to zero, as the following transformation equations clarify: \enl

\noindent
{\em Transverse $v = \pm\,1$ modes\/}:
\bea
(\tl{E}_{-}\mp \tl{H}_{1}) & = & (E_{-}\mp H_{1})\,
(\cos^{2}\varphi-\sin^{2}\varphi)
- 2\,(E_{1}\pm H_{-})\,\cos\varphi\,\sin\varphi \\
(\tl{E}_{1}\pm \tl{H}_{-}) & = & (E_{1}\pm H_{-})\,
(\cos^{2}\varphi-\sin^{2}\varphi)
+ 2\,(E_{-}\mp H_{1})\,\cos\varphi\,\sin\varphi \ ;
\eea
hence, the magnitude of the difference between the two angles that
set one or the other polarisation state of the transverse component
to zero is given by $|\varphi_{1}-\varphi_{2}| =
\sfrac{\pi}{4}$. \enl

\noindent
{\em Longitudinal $v = \pm\,\sfrac{1}{2}$ modes\/}:
\bea
(\tl{E}_{3}\mp \tl{H}_{2}) & = & (E_{-}\mp H_{1})\,\cos\varphi
- (E_{2}\pm H_{3})\,\sin\varphi \\
(\tl{E}_{2}\pm \tl{H}_{3}) & = & (E_{2}\pm H_{3})\,\cos\varphi
+ (E_{3}\mp H_{2})\,\sin\varphi \ ;
\eea
in this case, the magnitude of the difference between the two
angles that set one or the other polarisation state of the
longitudinal component to zero is now given by
$|\varphi_{1}-\varphi_{2}| = \sfrac{\pi}{2}$. \enl

Concluding the current considerations on the features of the
characteristic eigenfields of the Weyl curvature, for comparison we
can construct the complex scalars introduced by Newman and Penrose
\ct{newpen62} in the null tetrad canonically associated with our
$1+3$ ONF. With the help of Eqs. (C.26) and (C.27) given in
Ref. \ct{hve96}, we find (\,in the language of Szekeres
\ct{sze65,sze66}\,) for the incoming and outgoing transverse
radiative parts,
\be
\Psi_{0} = \sfrac{1}{\sqrt{3}}\,(E_{-}+H_{1})
- i\,\sfrac{1}{\sqrt{3}}\,(E_{1}-H_{-}) \hsp5 \mbox{and} \hsp5
\Psi_{4} = \sfrac{1}{\sqrt{3}}\,(E_{-}-H_{1})
+ i\,\sfrac{1}{\sqrt{3}}\,(E_{1}+H_{-}) \ ,
\ee
for the Coulomb part
\be
\Psi_{2} = -\,\sfrac{1}{3}\,E_{+} - i\,\sfrac{1}{3}\,H_{+} \ ,
\ee
and for the incoming and outgoing longitudinal radiative parts
\be
\Psi_{1} = -\,\sfrac{1}{2\sqrt{3}}\, (E_{3}+H_{2})
+ i\,\sfrac{1}{2\sqrt{3}}\,(E_{2}-H_{3}) \hsp5 \mbox{and} \hsp5
\Psi_{3} = \sfrac{1}{2\sqrt{3}}\,(E_{3}-H_{2})
+ i\,\sfrac{1}{2\sqrt{3}}\,(E_{2}+H_{3}) \ .
\ee
%

\subsection{Propagation of constraints}
The analysis of the propagation of the constraints (\r{onfdivsig})
- (\r{onffried}) is a slightly delicate issue, as in the given
context the tangents ${\bf u}$ to a generic matter flow have
non-zero vorticity; hence, there exist no well-defined spacelike
3-surfaces ${\cal T}$: $\left\{t=\mbox{const}\right\}$ everywhere
orthogonal to ${\bf u}$, and the frame derivatives $\p_{\alpha}$
contain a temporal partial derivative according to
Eq. (\r{onf13}). In principle this problem could be circumvented by
slicing $\left(\,{\cal M},\,{\bf g},\,{\bf u}\,\right)$ with any
arbitrary family of spacelike 3-surfaces ${\cal T}$:
$\left\{t=\mbox{const}\right\}$, on which one alternatively
evaluates the constraints and propagates them along the timelike
normals (which will be tilted with respect to ${\bf u}$). In this
non-comoving approach the energy-momentum-stress tensor would
appear imperfect. A preliminary investigation has revealed that
direct propagation {\em along\/} ${\bf u}$ of the subset
$(C_{1})^{\alpha}$ to $(C_{5})^{\alpha}$ in $1+3$ covariant terms
preserves these conditions \ct{hve98}.

Recently it has been shown by Maartens \ct{maa97} and Velden
\ct{vel97} that in the subcase when $p = 0$ and $0 = \udot^{\alpha}
= \om^{\alpha}$ (`irrotational dust') all constraints are preserved
along ${\bf u}$ and unique solutions exist if the initial data are
analytic (\,see also Ref. \ct{mac98}\,). It should be possible to
extend this to the non-analytic case by enlarging the FOSH set to
include the evolution equations for the constraints as
well. Indeed, Friedrich derives and discusses a FOSH evolution
system for the propagation of constraints of a particular FOSH
reduction of the vacuum EFE based on a combination of ADM $3+1$ and
Bianchi field equations \ct{fri96}. We anticipate that extension to
at least an irrotational perfect fluid should be
straightforward. We explicitly demonstrate this below in the
particular case of cosmological models in LRS class II.

\subsection{Propagation of inhomogeneities}
It is instructive to take a qualitative look at how, in
cosmological models $\left(\,{\cal M},\,{\bf g},\,{\bf u}\,\right)$
with barotropic perfect fluid matters source fields, changes in the
distribution of the matter are communicated from one spacetime
event to a nearby one. For this reason let us define the $1+3$
covariant gradient variables $X_{a} := \D_{a}\mu$ and $Z_{a} :=
\D_{a}\Th$. It is then fairly straightforward to derive the
following set of evolution equations (\,see, e.g.,
Refs. \ct{ellbru89} and \ct{hve96}\,)
\bea
\l{xdot}
\dot{X}{}^{\lgl a\rgl} & = & -\,\sfrac{4}{3}\,\Th\,X^{a}
- \sig^{a}\!_{b}\,X^{b} + \eps^{abc}\,\om_{b}\,X_{c}
- (\mu+p)\,Z^{a}
\\ \nonumber \\
\l{zdot}
c_{s}^{2}\,\dot{Z}{}^{\lgl a\rgl}
- c_{s}^{2}\,\D^{a}(\D_{b}\udot^{b}) & = & c_{s}^{2}\,
\D^{a}[\ (\udot_{b}\udot^{b}) - (\sig^{b}\!_{c}\sig^{c}\!_{b})
+ 2(\om_{b}\om^{b})\ ] - c_{s}^{2}\,[\ \Th\,Z^{a} + \sig^{a}\!_{b}
\,Z^{b} - \eps^{abc}\,\om_{b}\,Z_{c} + \sfrac{1}{2}\,X^{a}\ ]
\nonumber \\
& & \hsp5 + \ c_{s}^{2}\,\udot^{a}\ [\ (\D_{b}\udot^{b})
+ (\udot_{b}\udot^{b}) - \sfrac{1}{3}\,\Th^{2}
- (\sig^{b}\!_{c}\sig^{c}\!_{b}) + 2(\om_{b}\om^{b}) + \mu\ ]
\\ \nonumber \\
\l{divudotdot}
[\ (\D_{a}\udot^{a})\ ]\,\dot{} - c_{s}^{2}\,\D_{a}Z^{a}
& = & -\,[\ c_{s}^{-2}\,\frac{d^{2}p}{d\mu^{2}}\,(\mu+p)
- c_{s}^{2} + \sfrac{2}{3}\ ]\ \Th\ (\D_{a}\udot^{a})
- 2\,\sig_{ab}\,[\ \D^{\lgl a}\udot^{b\rgl} + \udot^{\lgl a}\,
\udot^{b\rgl}\ ] + 2\,\udot_{a}\,\eps^{abc}\,\D_{b}\om_{c}
\nonumber \\
& & \hsp5 - \ \frac{d^{2}p}{d\mu^{2}}\,(\mu+p)\,\Th\,\udot^{a}\,
\D_{a}c_{s}^{-2} - 2\,[\ c_{s}^{-2}\,\frac{d^{2}p}{d\mu^{2}}\,
(\mu+p) - c_{s}^{2} + \sfrac{1}{2}\ ]\,\udot_{a}\,Z^{a}
\nonumber \\
& & \hsp5 + \ [\ c_{s}^{-2}\,\frac{d^{3}p}{d\mu^{3}}\,(\mu+p)
+ (c_{s}^{-2}-1)\,\frac{d^{2}p}{d\mu^{2}}\ ]\,c_{s}^{-2}\,(\mu+p)\,
\Th\,(\udot_{a}\udot^{a}) + (c_{s}^{2}+\sfrac{1}{3})\,\Th\,
(\udot_{a}\udot^{a}) \ .
\eea
As noticed above from Eq. (\r{chef1}) for the spacetime metric
${\bf g}$ and the matter energy density $\mu$, we now find that
also the (magnitude of the locally) spatial gradient of $\mu$
propagates along the integral curves of ${\bf u}$, that is, growth
or decay of $X_{a}$ is entirely determined locally.

Where then do the sound waves occur in this first-order formulation
of the equations? In fact, information on inhomogeneities in the
matter distribution is propagated through a dynamical coupling
(determined by the local equation of state) of the spatial gradient
of the fluid expansion to the spatial divergence of the fluid
acceleration, enabling waves travelling at the speed of sound which
then pass on information to further fluid kinematical variables
and, hence, by the kinematical interaction terms on the right-hand
side of Eq. (\r{xdot}), to the density. When $c_{s} = 0$,
information can only be propagated through the Weyl
curvature. Subcases for which even this vehicle of communication is
unavailable (due to dynamical restrictions on the initial data)
have been called `silent' cosmological models
\ct{matetal94,hveetal97}; in them, {\em all\/} initial data are
merely evolved along ${\bf u}$.

\subsection{Implications of the $v = \pm\,\sfrac{1}{2}$ Weyl
curvature characteristic eigenfields}
The key issue now is what to make of the characteristics along
which the longitudinal components of the Weyl curvature travel at
speed $|v| = \sfrac{1}{2}$ relative to observers comoving with
${\bf u}$, the latter being physically and geometrically uniquely
defined in the perfect fluid case we consider here. They appear to
represent the possibility of propagation of longitudinal
gravitational waves at (in the given frame) one half the speed of
light, thus corresponding in some sense to massive rather than
massless particles. Is this really so, or is there some way to
avoid this conclusion? We consider here a series of relevant
issues; our conclusion is that none of them is at present
sufficient to exclude such longitudinal gravitational waves.

\subsubsection{Can they be gauged away?}
The Weyl curvature itself is of course an invariant geometrical
quantity (and is gauge-invariant for linearised gravitational
perturbations of Minkowski spacetime or FLRW cosmologies).
However, various Weyl curvature frame components can be set to zero
if one is allowed {\em arbitrary\/} choice of the reference frame
--- see, for example, the discussions on the Petrov classification
of the Weyl curvature of (vacuum) gravitational fields in the null
tetrad literature, in which one of the normal forms for the
algebraically general case sets the longitudinal components to zero
\ct{ksmh80}, but another instead sets the transverse components to
zero \ct{cha83}. But that frame freedom is not allowed in the
situation considered here, where $\{\,\p_{a}\,\}$ is tied in to the
Fermi-propagated reference frame of fundamental observers --- which
is the relevant frame to consider in the context of detection of
gravitational radiation via the GDE \ct{pir56}. Thus, as emphasized
in the previous section, one can use the remaining frame freedom to
set one of the longitudinal polarisation modes to zero, but not to
remove all components of a propagating longitudinal gravitational
wave. There is not enough frame freedom remaining.

Furthermore, these modes do indeed cause physically detectable
effects, via the GDE (cf. the discussion in the introduction, and
the display of the GDE source terms due to the longitudinal waves
that is given in appendix \r{app5}). In principle, one could cancel
these effects by accelerating and rotating the detector in
precisely the right way, if one knew in advance what information
was going to be contained in the wave\footnote{Charles Hellaby
(private communication) suggested a nice analogy: in principle one
can shake one's head in such a way as to not hear a sound wave.}
--- but that does not correspond to motion of a physically
realistic detector; and if we were somehow to achieve this, the
information contained in these wave modes would then be coded in
the requisite motion of the detector (which could only be achieved
by use of suitable rocket motors or other energy-consuming physical
actuators). Information (and energy) would indeed have been
conveyed in a physically detectable fashion from the source to the
observer.

\subsubsection{Does some consistency condition prevent them
occurring?}
We have been unable to see any reason why solution of the initial
constraint equations should prevent such modes appearing. Taking
linear combinations of Eqs. (\r{onfdive}) and (\r{onfdivh}) one
finds that the propagation direction adapted, divergence-like
$\p_{1}$ frame derivative of the transverse modes is unconstrained
whereas that of the longitudinal modes has typically non-zero
sources. Further, we cannot see why consistency conditions between
the time evolution and constraint equations should forbid them. To
explore this a bit further, we have looked at the form of the
equations when linearised about a FLRW model; the Weyl curvature
variables for this case are given in appendix \r{app4}, the related
exact propagation equations in appendix \r{app2}. There seems to be
no barrier to the linearised equations having such solutions.

\subsubsection{Does some conservation law prevent their emission?}
Overall conservation of mass and momentum, resulting in asymptotic
conservation laws in asymptotically flat space-times (\,see, e.g.,
Bondi et al \ct{bonetal62}\,), are global conservation laws over
the entire sky. It is not clear how they could prevent emission of
radiation of the type identified here in specific directions,
provided the overall balance over the whole sky was achieved.

To properly examine conservation laws associated with this kind of
radiation, one would, however, have to rework the previous
asymptotic analysis in terms of coordinates adapted to the
semi-null cone corresponding to a speed of propagation $|v| =
\sfrac{1}{2}$, rather than the Lorentz-invariant null cone
itself. We have not attempted that extension of previous work, but
see no reason to believe it would lead to conservation laws that
would forbid radiation such as identified here.

\subsubsection{Why have they not been detected in previous work?}
Given that these longitudinal modes seem to be unequivocally
implied by the FOSH form of the $1+3$ covariant equations, why have
they not been identified before?  Particularly, for example, why do
they not appear in the somewhat similar analysis of Ehlers et al
\ct{ehletal87}, examining propagation of linearised perturbations
of a `dust' spacetime geometry?

The answer seems to be that they have not been looked for, in that
in most such studies the characteristics of propagation of the Weyl
curvature were not examined; rather, those of the metric and its
first derivatives were the focus of attention. In that case, one
will not necessarily detect these modes, for the metric itself has
as its characteristics the timelike coordinate lines associated
with the fundamental observers (\,cf. also
Ref. \ct{putear96}\,). The key point is that, given a second-order
hyperbolic equation for some field, {\em the 3-surfaces across
which discontinuities of the field and/or its first derivatives can
occur are not necessarily the same as those across which
discontinuities of its second derivatives can occur\/}. It is the
latter that we have been investigating here, because that is the
level at which invariant physical effects occur.

The situation may actually be analagous to that pointed out above
for the matter fluid energy density, which does not itself
immediately manifest the sound cone characteristics; these affect
the density through first being manifest in other variables which
then affect the density itself (\,by appearing in the kinematical
interaction terms on the right-hand side of Eq. (\r{xdot})\,).

Indeed, when attention has been placed on the Weyl curvature
itself, these modes have been discovered before by Friedrich, who
examined FOSH evolution systems for vacuum gravitational fields
that employed the Bianchi field equations \ct{fri96}.\footnote{He
does not mention these modes in his fluid paper \ct{fri98}; but
they must be there in the system he has examined in that paper, for
they were there in the vacuum case.} However, he discounted them
because the related characteristic 3-surfaces of propagation were
not Lorentz-invariant, and because an invariantly defined timelike
vector field did not exist.\footnote{Helmut Friedrich (private
communication).} As indicated above, we do not believe this is a
good enough reason in our barotropic perfect fluid case to claim
these are not physical modes that can transport energy from one
place to another. Our emphasis on the physics associated with the
GDE indicates the contrary. As already remarked by Szekeres
\ct{sze65,sze66}, the longitudinal wave component of the Weyl
curvature does contribute to the GDE force term, to give measurable
effects. The relevant GDE terms are given for both null and
timelike congruences in appendix \r{app5}.

Interestingly, special propagation processes of disturbances in
gravitational fields along {\em non-null\/} curves were described
before by Araujo in the context of non-linear interactions between
three gravitational waves in a vacuum spacetime geometry
\ct{ara89}.

\subsubsection{How can they have a propagation speed that is
not Lorentz-invariant?}
In the specific case we consider, Lorentz invariance is broken by
the existence of a cosmological fluid with a preferred 4-velocity
vector field ${\bf u}$. Many of the variables we have identified,
including the Coulomb-like components of the Weyl curvature, have
the associated timelike curves with $v = 0$ as (degenerate)
characteristics; and these are certainly not Lorentz-invariant, but
we are not surprised by their existence. Similarly, also the $v =
\pm\,\sfrac{1}{2}$ modes are not Lorentz-invariant. When boosting
to a reference frame that does {\em not\/} comove with ${\bf u}$,
in the perfect fluid case the longitudinal Weyl curvature modes
typically get mixed up with the matter fluid variables $\mu$ and
$p$ and the characteristic structure as well as all propagation
speeds other than $|v| = 1$ will look different in this
frame.\footnote{The given $v = 0$ and $v = \pm\,c_{s}$ modes are
not Lorentz-invariant either; however, in the generic case there
are, e.g., Coulomb-like components of the Weyl curvature with $|v|
= 0$ in {\em any\/} rest frame.} However, we determine propagation
velocities with respect to a specific set of physical observers,
and {\em this\/} breaks Lorentz invariance. (For completeness, we
give the independent Lorentz-invariant scalars one can construct
from the Weyl curvature tensor in appendix \r{app3}.)

While we are surprised by what is indicated by the equations, the
non-Lorentz-invariant nature of the characteristics for the
longitudinal modes does not by itself mean they do not occur. On
the contrary, if the standard theory of FOSH evolution systems is
indeed reliable, these are characteristics along which
discontinuities and waves can propagate. The possibility of
existence of associated physical waves has to be taken seriously
unless some reason is found why they cannot exist. As indicated
above, we are unaware of any such reason.

\section{Worked example: Cosmological models in LRS class II}
\l{sec:foshlrs2}
After derivation of a 44-D FOSH evolution system for generic
cosmological models $\left(\,{\cal M},\,{\bf g},\,{\bf u}\,\right)$
with barotropic perfect fluid matter sources, based on geometrical
fields defined in the $1+3$ ONF formalism, we now turn to discuss
in some detail a relatively simple, lower-dimensional, subcase,
namely cosmological models in Locally Rotationally Symmetric
(`LRS') class II \ct{steell68,hveell96}. This class comprises
cosmologies of either spherical, plane or hyperbolic spacetime
symmetry that are typically spatially inhomogeneous. The first to
present a FOSH formulation for the {\em spherically symmetrical\/}
case were Kind and Ehlers \ct{kinehl93}. Their work, which, besides
the matter energy density, employed dynamical field variables
derived from metric functions and their derivatives, led to a 8-D
FOSH evolution system subject to five constraints, the dependent
field variables being (in their notation) $u^{A} = [\ X, \,Y, \,Q,
\,\Lambda, \,{\cal L}, \,w, \,\om, \,\mu\ ]^{T}$. The dynamical
formulation to follow yields a result of similar structure. It
represents a 8-D invariant subspace of our full 44-D FOSH evolution
system which employs geometrical variables. Ultimately, we will
explicitly include the conservation of the constraint equations to
build a larger FOSH evolution system. This latter aspect will lead
to a natural solution for the threading lapse function $M$ (which
we do {\em not\/} assume to be the constant $M_{0}$ of the previous
section). Besides these issues, our formulation shows clearly that
the intrinsic geometry of the 2-D spacelike symmetry group orbits
can be fully decoupled from the evolution of the physically
relevant field variables. Due to the spacetime symmetry imposed,
these models exclude all (forms of) gravitational radiation; thus,
apart from sound waves, these are gravitationally `silent'
universes, as mentioned above.

In the $1+3$ ONF formalism, LRS spacetime geometries are
characterised as follows \ct{hveugg97}: the unit tangent of the
invariantly defined preferred spacelike direction may be taken as,
e.g., $\p = \p_{1}$; then $0 = \p_{2}(f) = \p_{3}(f)$ for any
invariantly defined geometrical variable. Rank 2 symmetric
tracefree tensors ${\bf A}$ orthogonal to ${\bf u}$ specialise to
$0 = A_{-} = A_{1} = A_{2} = A_{3} \Rightarrow A^{2} =
\sfrac{1}{3}\,(A_{+})^{2} \Rightarrow A = \pm\,\sfrac{1}{\sqrt{3}}
\,A_{+}$ (\,cf. Eq. (\r{apm})\,). The case of our interest, the
cosmological models in LRS class II, are defined by the conditions
$0 = \om = n_{11} \Rightarrow H_{+} = 0$ \ct{steell68,hveugg97}.
Then it can be shown that the general relations listed in section
\r{sec:13eqs} reduce to (\,cf. Ref. \ct{hve96}\,) \enl

\noindent
{\em The commutators\/}:
\bea
\l{onflrs2com1}
\left[\,{\bf e}_{0}, {\bf e}_{1}\,\right] & = & \dot{u}\,
{\bf e}_{0} - \sfrac{1}{3}(\Th-2\sig_{+})\,{\bf e}_{1} \\
\l{onflrs2com2}
\left[\,{\bf e}_{0}, {\bf e}_{2}\,\right] & = & - \,\sfrac{1}{3}
(\Th+\sig_{+})\,{\bf e}_{2} \\
\l{onflrs2com3}
\left[\,{\bf e}_{0}, {\bf e}_{3}\,\right] & = & - \,\sfrac{1}{3}
(\Th+\sig_{+})\,{\bf e}_{3} \\
\l{onflrs2com4}
\left[\,{\bf e}_{1}, {\bf e}_{2}\,\right] & = & a\,{\bf e}_{2} \\
\l{onflrs2com5}
\left[\,{\bf e}_{2}, {\bf e}_{3}\,\right] & = & 2\,n_{31}\,
{\bf e}_{3} \\
\l{onflrs2com6}
\left[\,{\bf e}_{3}, {\bf e}_{1}\,\right] & = & - \,a\,{\bf e}_{3}
\ .
\eea

\noindent
{\em The evolution equations\/}: \nopagebreak
\bea
\l{onflrs2thdot}
{\bf e}_{0}(\Th) & = & - \,\sfrac{1}{3}\,\Th^{2}
+ ({\bf e}_{1} + \dot{u} - 2a)\,(\dot{u}) -
\sfrac{2}{3}\,(\sig_{+})^{2} - \sfrac{1}{2}\,(\mu+3p) \\
\l{onflrs2sigdot}
{\bf e}_{0}(\sig_{+}) & = & - \,\sfrac{1}{3}\,(2\Th-\sig_{+})
\,\sig_{+} - ({\bf e}_{1}+\dot{u}+a)\,(\dot{u}) - E_{+} \\
\l{onflrs2udotdot}
{\bf e}_{0}(\dot{u}) & = & (c_{s}^{2}-\sfrac{1}{3})\,\Th\,\dot{u}
+ c_{s}^{2}\,{\bf e}_{1}(\Th) + \Th\,\frac{d^{2}p}{d\mu^{2}}
\,{\bf e}_{1}(\mu) + \sfrac{2}{3}\,\sig_{+}\,\dot{u} \\
\l{onflrs2adot}
{\bf e}_{0}(a) & = & - \,\sfrac{1}{3}(\Th+\sig_{+})\,
(\dot{u}+a) \\
\l{onflrs2n31dot}
{\bf e}_{0}(n_{31}) & = & - \,\sfrac{1}{3}(\Th+\sig_{+})\,
n_{31} \\
\l{onflrs2edot}
{\bf e}_{0}(E_{+}) & = & - \,\sfrac{1}{2}\,(\mu+p)\,\sig_{+}
- (\Th+\sig_{+})\,E_{+} \\
\l{onflrs2mudot}
{\bf e}_{0}(\mu) & = & - \,\Th\,(\mu+p) \ .
\eea

\noindent
{\em The constraint equations\/}:
\bea
\l{onflrs2divsig}
0 & = & (C_{1}) \ := \ ({\bf e}_{1}-3a)\,(\sig_{+})
+ {\bf e}_{1}(\Th) \\
\l{onflrs2ajac}
0 & = & (C_{2}) \ := \ ({\bf e}_{1}-a)\,(a) + \sfrac{1}{9}\,\Th^{2}
- \sfrac{1}{9}\,(\Th+2\sig_{+})\,\sig_{+} + \sfrac{1}{3}\,E_{+}
- \sfrac{1}{3}\,\mu \\
\l{onflrs2n31jac}
0 & = & (C_{3}) \ := \ ({\bf e}_{1}-a)\,(n_{31}) \\
\l{onflrs2dive}
0 & = & (C_{4}) \ := \ ({\bf e}_{1}-3a)\,(E_{+})
+ \sfrac{1}{2}\,{\bf e}_{1}(\mu) \\
\l{onflrs2mom}
0 & = & (C_{PF}) \ := \ c_{s}^{2}\,{\bf e}_{1}(\mu)
+ (\mu+p)\,\dot{u} \ .
\eea
Note that Eqs. (\r{onflrs2n31dot}) and (\r{onflrs2n31jac}) for the
spatial commutation function $n_{31}$ decouple from the remaining
set of equations; $n_{31}$ in fact determines the \enl

\noindent
{\em Gau\ss ian curvature of 2-D spacelike symmetry group
orbits\/}: \nopagebreak
\be
\l{onflrs2gk}
K := 2\,({\bf e}_{2}-2n_{31})\,(n_{31}) \ ;
\ee
the group orbits can either be spherically symmetrical ($K > 0$),
plane symmetrical ($K = 0$), or hyperbolically symmetrical ($K <
0$). \enl

\noindent
{\em Tracefree part and trace of 3-Ricci curvature\/}:
\bea
\l{onflrs3ric1}
{}^{*}\!S_{+} & = & -\,{\bf e}_{1}(a) + K
\ = \ E_{+} - \sfrac{1}{3}(\Th+\sig_{+})\,\sig_{+}
+ (C_{G})_{+} \\
\l{onflrs3rscl}
^{*}\!R & = & 2\,(2{\bf e}_{1}-3a)\,(a) + 2\,K
\ = \ 2\,\mu - \sfrac{2}{3}\,\Th^{2} + \sfrac{2}{3}\,
(\sig_{+})^{2} + (C_{G}) \ .
\eea
Combining Eqs. (\r{onflrs2ajac}) and (\r{onflrs3ric1}), one obtains
for $K$ the algebraic expression
\be
K = \sfrac{2}{3}\,E_{+} + \sfrac{1}{3}\,\mu - \sfrac{1}{9}\,
(\Th+\sig_{+})^{2} + a^{2} + (C_{G})_{+} + (C_{2}) \ .
\ee
%

\subsection{Set of FOSH evolution equations}
We choose our set of (dimensionless) comoving local coordinates as
$\{\,x^{\mu}\,\} = \{\,t, \,x, \,y, \,z\,\}$. Since $\om^{\alpha} =
0 \Rightarrow M_{i} = 0$, the expansion of the frame basis fields
$\p_{a}$ of Eq. (\r{onf13}) reduces to a zero-shift-vector version
(`Eulerian observers') of the ADM $3+1$ slicing formulation of the
dynamics of relativistic spacetime geometries \ct{adm62}; with
properly defined spacelike 3-surfaces ${\cal T}$:
$\left\{t=\mbox{const}\right\}$. For LRS class II we find in
particular \ct{hve96,steell68}
\bea
\l{onflrs231}
{\bf e}_{0} = M^{-1}\,\partial_{t} \ , \hsp5
{\bf e}_{2} = Y^{-1}\,\partial_{y} \ , \hsp5
{\bf e}_{1} = X^{-1}\,\partial_{x} \ , \hsp5
{\bf e}_{3} = (YZ)^{-1}\,\partial_{z} \ ,
\eea
where $M = M(t,x)$, $X = X(t,x)$, $Y = Y(t,x)$ and $Z = Z(y)$; each
of the $\p_{a}$ here is hypersurface orthogonal. With the
simplifications due to the spacetime symmetry of LRS class II
given, our set of dynamical field variables is comprised of the 8-D
subset $u^{A} = [\,\udot, \,\sfrac{1}{3}(\Th-2\sig_{+}),
\,\sfrac{1}{3}(\Th+\sig_{+}), \,a, \,E_{+}, \,\mu, \,X, \,Y\,]^{T}$
of Eq. (\r{depvar}), leading to the FOSH evolution system
\bea
\l{fosh1}
M^{-1}\,\ptl_{t}\udot - c_{s}^{2}\,X^{-1}\,
\ptl_{x}\sfrac{1}{3}(\Th-2\sig_{+})
& = & -\,[\ c_{s}^{-2}\,\frac{d^{2}p}{d\mu^{2}}\,
(\mu+p)-c_{s}^{2} + \sfrac{1}{3}\ ]\ \Th\,\udot + \sfrac{2}{3}\,
\sig_{+}\,\udot + 2\,c_{s}^{2}\,a\,\sig_{+} \nonumber \\
& & \hspace{20mm} - \ c_{s}^{2}\,(C_{1}) + \Th\,c_{s}^{-2}\,
\frac{d^{2}p}{d\mu^{2}}\,(C_{PF})
\\ \nonumber \\
\l{fosh2}
c_{s}^{2}\,M^{-1}\,\ptl_{t}\sfrac{1}{3}(\Th-2\sig_{+})
- c_{s}^{2}\,X^{-1}\,\ptl_{x}\udot 
& = & -\,c_{s}^{2}\,[\ \sfrac{1}{9}\,(\Th-2\sig_{+})^{2}
- \udot^{2} + \sfrac{1}{6}\,(\mu+3p) - \sfrac{2}{3}\,
E_{+}\ ]
\\ \nonumber \\
\l{fosh3}
M^{-1}\,\ptl_{t}\sfrac{1}{3}(\Th+\sig_{+}) & = &
-\,\sfrac{1}{9}\,(\Th+\sig_{+})^{2} - a\,\udot
- \sfrac{1}{6}\,(\mu+3p) - \sfrac{1}{3}\,E_{+}
\\ \nonumber \\
\l{fosh4}
M^{-1}\,\ptl_{t}a & = & -\,\sfrac{1}{3}(\Th+\sig_{+})\,(\udot+a)
\\ \nonumber \\
\l{fosh5}
M^{-1}\,\ptl_{t}E_{+} & = & -\,\sfrac{1}{2}\,(\mu+p)\,\sig_{+}
- (\Th+\sig_{+})\,E_{+}
\\ \nonumber \\
\l{fosh6}
M^{-1}\,\ptl_{t}\mu & = & -\,\Th\,(\mu+p)
\\ \nonumber \\
\l{fosh7}
M^{-1}\,\ptl_{t}X & = & \sfrac{1}{3}(\Th-2\sig_{+})\,X
\\ \nonumber \\
\l{fosh8}
M^{-1}\,\ptl_{t}Y & = & \sfrac{1}{3}(\Th+\sig_{+})\,Y \ .
\eea
We note in passing that, like $n_{31}$ in Eq. (\r{onflrs2n31dot}),
the evolution of the variable $Y$, which will be found to be
equivalent to the Gau\ss ian curvature according to $K =
C_{1}/Y^{2}$, is decoupled from the remaining equations; thus the
resulting set is effectively 7-D.

The symmetric matrices $M^{AB\,\mu}$ entering in Eq. (\r{fosh})
assume in the given situation the explicit forms
\bea
\l{mabmu}
M^{AB\,0} =
\left( \begin{array}{cccccccc}
        1 & .         & . & . & . & . & . & . \\
        . & c_{s}^{2} & . & . & . & . & . & . \\
        . & .         & 1 & . & . & . & . & . \\
        . & .         & . & 1 & . & . & . & . \\
        . & .         & . & . & 1 & . & . & . \\
        . & .         & . & . & . & 1 & . & . \\
        . & .         & . & . & . & . & 1 & . \\
        . & .         & . & . & . & . & . & 1
       \end{array}
\right) \ , \hsp5
M^{AB\,1} = -\,M\,X^{-1}\,c_{s}^{2}
\left( \begin{array}{cccccccc}
        . & 1 & . & . & . & . & . & . \\
        1 & . & . & . & . & . & . & . \\
        . & . & . & . & . & . & . & . \\
        . & . & . & . & . & . & . & . \\
        . & . & . & . & . & . & . & . \\
        . & . & . & . & . & . & . & . \\
        . & . & . & . & . & . & . & . \\
        . & . & . & . & . & . & . & .
       \end{array}
\right) \ .
\eea
The FOSH evolution system (\r{fosh1}) - (\r{fosh8}) will be fully
deterministic, if either we prescribe the threading lapse function
$M$ by hand, or we have a further equation for its evolution along
${\bf u}$. We will return to this issue further below.

The characteristic condition $0 =
\det\,[\,M^{AB\,\mu}\,\xi_{\mu}\,]$, with, in the present case,
$\xi_{\mu} = v\,M\,\d^{0}\!_{\mu} + X\,\d^{1}\!_{\mu}$, reads \enl

\noindent
{\em The characteristic condition\/}:
\be
0 = \det\,[\,v\,M\,M^{AB\,0} + X\,M^{AB\,1}\,]
= M^{8}\,c_{s}^{2}\,v^{6}\,(v-c_{s})\,(v+c_{s}) \ .
\ee
Hence, the characteristic propagation velocities with respect to an
observer at rest in $\{\,\p_{a}\,\}$ are either $v_{1} = 0$ or
$v_{2,3} = \pm\,c_{s}$. Upon diagonalisation of the principle part
of Eqs. (\r{fosh1}) - (\r{fosh8}), the characteristic eigenfields
propagating along the sound cone are given by
\be
u_{(2)}^{A} = [\ \sfrac{1}{\sqrt{1+c_{s}^{2}}}\,
(\,c_{s}\udot\pm\sfrac{1}{3}(\Th-2\sig_{+})\,)\ ]^{T} \ ,
\ee
a result which, as to be expected, is just a specialisation of the
general one obtained for barotropic perfect fluids in
Eq. (\r{chef2}) of section \r{sec:foshpf}.

Setting $p = 0 \Rightarrow \udot = 0$, $c_{s} = 0$ (for $K > 0$),
we get from Eqs. (\r{fosh1}) - (\r{fosh8}) the evolution part that
describes the (`silent', and spherically symmetrical)
Lema\^{\i}tre--Tolman--Bondi cosmological models.

\subsection{Propagation of constraints}
Using the coordinate components of $\{\,\p_{a}\,\}$ given in
Eq. (\r{onflrs231}), the commutator relations (\r{onflrs2com1}),
(\r{onflrs2com4}) and (\r{onflrs2com5}) yield
\bea
\l{cudot}
0 & = & (C_{\udot}) \ := \ \udot - (MX)^{-1}\,\ptl_{x}N \\
\l{ca}
0 & = & (C_{a}) \ := \ a + (YX)^{-1}\,\ptl_{x}Y \\
\l{cn31}
0 & = & (C_{n_{31}}) \ := \ n_{31} + \sfrac{1}{2}\,(ZY)^{-1}\,
\ptl_{y}Z \ ,
\eea
respectively. Together with Eq. (\r{cn31}), Eq. (\r{onflrs2gk})
integrates to $K = C_{1}/Y^{2}$, where $C_{1}$ denotes a
dimensionless integration constant. One can show that the set of
constraints comprised of Eqs. (\r{onflrs2divsig}) -
(\r{onflrs2mom}) and (\r{cudot}) - (\r{cn31}) propagates according
to
\bea
\l{c1dot}
M^{-1}\,\ptl_{t}(C_{1}) & = & -\,\Th\,(C_{1}) - 3\,\udot\,(C_{2})
- (C_{4}) - \sfrac{3}{2}\,(C_{PF}) \nonumber \\
& & \hspace{20mm} + \ [\ \sfrac{1}{3}\,\Th^{2}+\sfrac{2}{3}\,\Th
\sig_{+}+\sfrac{1}{3}(\sig_{+})^{2}+3\,a\,\udot+\sfrac{1}{2}\,
(\mu+3p)+E_{+}\ ]\ (C_{\udot})
\\ \nonumber \\
M^{-1}\,\ptl_{t}(C_{2}) & = & -\,\sfrac{1}{3}\,(2\Th-\sig_{+})\,
(C_{2}) - \sfrac{1}{3}\,(\udot+a)\,(C_{1})
+ \sfrac{1}{3}(\Th+\sig_{+})\,(\udot+a)\,(C_{\udot})
\\ \nonumber \\
M^{-1}\,\ptl_{t}(C_{3}) & = & -\,\sfrac{1}{3}\,(2\Th-\sig_{+})\,
(C_{3}) - \sfrac{1}{3}\,n_{31}\,(C_{1})
+ \sfrac{1}{3}(\Th+\sig_{+})\,n_{31}\,(C_{\udot})
\\ \nonumber \\
M^{-1}\,\ptl_{t}(C_{4}) & = & -\,\sfrac{1}{3}\,(4\Th+\sig_{+})\,
(C_{4}) - \sfrac{1}{2}\,(\mu+p)\,(C_{1}) - E_{+}\,(C_{1})
- \sfrac{1}{2}\,(\Th+\sig_{+})\,(C_{PF}) \nonumber \\
& & \hspace{20mm} + \ \sfrac{1}{2}\,(\Th+\sig_{+})\,(\mu+p)\,
(C_{\udot}) + (\Th+\sig_{+})\,E_{+}\,(C_{\udot})
\\ \nonumber \\
M^{-1}\,\ptl_{t}(C_{PF}) & = & -\,[\ (c_{s}^{2}+\sfrac{4}{3})\,\Th
+ \sfrac{2}{3}\,\sig_{+}\ ]\,(C_{PF}) + c_{s}^{2}\,\Th\,
(\mu+p)\,(C_{\udot})
\\ \nonumber \\
\l{cudotdot}
M^{-1}\,\ptl_{t}(C_{\udot}) & = &
-\,\sfrac{1}{3}(\Th-2\sig_{+})\,(C_{\udot})
- \left[\ (MX)^{-1}\,\ptl_{t}(M^{-1}\ptl_{x}M)
- X^{-1}\,\ptl_{x}(c_{s}^{2}\Th) - c_{s}^{2}\,\Th\,\udot\ \right]
\\ \nonumber \\
M^{-1}\,\ptl_{t}(C_{a}) & = & -\,\sfrac{1}{3}(\Th-2\sig_{+})\,
(C_{a}) + \sfrac{1}{3}\,(C_{1}) - \sfrac{1}{3}(\Th+\sig_{+})\,
(C_{\udot})
\\ \nonumber \\
\l{cn31dot}
M^{-1}\,\ptl_{t}(C_{n_{31}}) & = &
-\,\sfrac{1}{3}(\Th+\sig_{+})\,(C_{n_{31}}) \ .
\eea
The constraints will propagate consistently, in fact also via a
FOSH evolution system (with {\em zero\/} characteristic propagation
velocites only), iff the evolution of the threading lapse function
$M$ (\,i.e., $\ptl_{t}M$\,) satisfies the relation
\be
\l{ndercon}
\ptl_{t}[\,M^{-1}\,\ptl_{x}M\,] = M\,\ptl_{x}[\,c_{s}^{2}\,\Th\,]
+ M\,X\,c_{s}^{2}\,\Th\,\udot \ ,
\ee
which follows from Eq. (\r{cudotdot}). That is, the evolution of
$M$ is determined through an integrability condition demanded by
preservation of the constraints. It turns out that the choice
(\r{threadm}) of Salzman and Taub satisfies Eq. (\r{ndercon})
subject to the validity of the constraint (\r{cudot}). The solution
(\r{threadm}) implies
\be
\l{ndot}
M^{-1}\,\ptl_{t}M = c_{s}^{2}\,\Th\,M \ .
\ee

Given the results of this subsection, it is natural to add
Eqs. (\r{c1dot}) - (\r{cn31dot}) to the set (\r{fosh1}) -
(\r{fosh8}) to form a larger FOSH evolution system for the set of
dynamical fields
\be
\tilde{u}{}^{A} = [\ \sfrac{1}{\sqrt{1+c_{s}^{2}}}\,
(\,c_{s}\udot\pm\sfrac{1}{3}(\Th-2\sig_{+})\,),
\,\sfrac{1}{3}(\Th+\sig_{+}), \,a, \,E_{+}, \,\mu, \,X, \,Y,
\,(C_{n})\ ]^{T} \ ,
\ee
which can be taken to model all spacetime geometries with perfect
fluid matter source fields in LRS class II.

\section{Conclusion}
The aim of the present work is to explore the propagational
features of physically relevant quantities in relativistic
cosmological models with idealised matter source fields that can be
treated as perfect fluids with barotropic equations of state. The
results we obtain on the characteristic structure of the derived
44-D FOSH evolution system apply to both, exact cases and
FLRW-linearised cosmological models; the principle part (the
dynamical interactions between various fields) remains unchanged
under the linearisation procedure, which affects the kinematical
interactions only. Hence, we can read off which kind of
perturbations propagate along which set of characteristic
3-surfaces in both the exact and the linearised theories. \enl

As to the question raised in Ref. \ct{hveetal97} concerning the
presence of gravitational radiation in spatially homogeneous but
anisotropic cosmological models with perfect fluid matter source
fields, it now becomes clear that, with respect to a
symmetry-group-invariant $1+3$ ONF $\{\,\p_{a}\,\}$, there then do
{\em not\/} exist any associated null characteristics (while they
are allowed by the dynamical equations, the homogeneous data does
not activate them). Only timelike ones are present, each with
propagation velocity $v = 0$ of the associated characteristic
eigenfields. Hence, there exists no {\em propagating\/}
gravitational radiation in such models (nor any other fields that
are exchanged between neighbouring worldlines of the matter
fluid). In this sence, this class of cosmological models can be
classified as `silent', too. For certain cases in this class it
may, however, be sensible to maintain some kind of interpretation
of the spacetime dynamics in terms of {\em standing\/}
gravitational waves (with respect to the timelike reference
congruence chosen); aspects along these lines have been discussed
by, e.g., Lukash \ct{luk75} or King \ct{kin91}. \enl

The $1+3$ ONF formulation with its geometrically defined dependent
field variables provides a possible alternative FOSH evolution
system for applications in numerical relativity, in particular to
relativistic cosmology (\,for a `living review' on the current
state of the latter see Ref. \ct{ann98}\,). Even though, in this
framework, the number of constraints, which are typically
non-linear elliptic partial differential equations, is larger than
in conventional schemes based on the ADM $3+1$ treatment, the
benefits may be substantial. Most variables in our treatment have
direct physical significance (e.g., the Weyl curvature and matter
variables enter the GDE); hence, they are intuitively more easily
accessible than the derivatives of metric components which appear
in various other works. The example of cosmological models in LRS
class II described in section \r{sec:foshlrs2} may serve as a
simple test bed for numerical simulations; analysis of
`shell-crossing singularities' in the Lema\^{\i}tre--Tolman--Bondi
`dust' case being one of possible problems to
investigate. Moreover, irrespective of the fact that a timelike
reference congruence ${\bf u}$ is {\em not\/} invariantly defined
for vacuum spacetime geometries, there is no reason that prevents
application of the $1+3$ ONF FOSH evolution system to those
situations as well, as recently emphasised by Friedrich
\ct{fri96}. For example, in Ref. \ct{schetal98} a 16-D FOSH
evolution system is presented for the numerical description of a
static Schwarzschild black hole in harmonic time slicing. Modifying
our general equations for $\mu = 0$ along the lines outlined for
spherically symmetrical cases in section \r{sec:foshlrs2} could
lead to a lower dimensional evolution system which may,
furthermore, not suffer from the continuum instabilities reported
in that article. There the authors attribute this problem to the
intrinsic structure of the equations they employed. That the FOSH
evolution system of the $1+3$ ONF formalism does {\em not\/} have
flux-conserving form may or may not have negative consequences in
prospective numerical investigations (\,cf. Ref. \ct{fri96} and the
`living review' \ct{reu98}\,). \enl

The most unexpected feature arising from our FOSH analysis of
invariantly defined quantities is the indication of the existence
of propagating longitudinal modes in the Weyl curvature. If this is
confirmed by further analysis, it has possible intriguing
implications in regard to (a) gravitational wave emission, (b)
gravitational wave detection, and (c) in relation to the
quantisation of the gravitational field. As regards the first two
issues, they presumably provide a further channel whereby energy
can be emitted from a source in the form of gravitational radiation
and then received in distant places (hence, resulting in detectable
effects). This would mean they need, for example, to be considered
seriously in the context of the inflationary universe scenarios for
gravitational wave emission and the associated CMBR
anisotropies. As regards the latter, they imply a substantial
difference relative to the vacuum electrodynamical case, where
longitudinal modes exist but have no physical effect and can be
gauged away.  However, in a plasma, where (charged) matter is
present as well as the electromagnetic field, the situation seems
rather similar to what is derived here.\footnote{Malcolm MacCallum
(private communication).}  We have shown that in the non-vacuum
gravitational case longitudinal modes appear that cannot be gauged
away; hence, possibly, these modes may need to be taken into
account when quantising gravitational radiation in some contexts.

\acknowledgments
We thank J\"{u}rgen Ehlers for stimulating discussions on FOSH
evolution systems which sparked the present investigation. Claes
Uggla helped us through numerous useful comments. We also
acknowledge helpful remarks by Marco Bruni and Malcolm
MacCallum. This work has been supported by the Deutsche
Forschungsgemeinschaft (DFG) and the South African Foundation for
Research and Development (FRD).

\appendix
\section*{}
\subsection{Conventions}
\l{app1}
Throughout our work we employ geometrised units which are defined
by setting $c = 1 = 8\pi G/c^{2}$. All geometrically defined
variables thus have physical dimensions that are integer powers of
$\lgth$ only. We denote spacetime indices with respect to an
arbitrary basis $\{\,\p_{a}\,\}$ by $a, \,b, \,c, \,\dots = 0 -
3$. When fixing the basis vectors to be orthonormal such that the
components of the metric tensor ${\bf g}$ reduce to the Minkowskian
form $g_{ab} = \eta_{ab} = \mbox{diag}\,(\,-1, \,1, \,1, \,1\,)$,
and aligning $\p_{0}$ with the future-directed, normalised tangents
${\bf u}$ of a congruence of preferred timelike curves, indices
with respect to the spatial basis $\{\,\p_{\alpha}\,\}$ are denoted
by $\alpha, \,\beta, \,\gam, \,\dots = 1 - 3$. For a set of
dimensionless local coordinates we write $\{\,x^{\mu}\,\} = \{\,t,
x^{i}\,\}$; spacetime and spatial coordinate indices being $\mu,
\,\nu, \,\rho, \,\dots = 0 - 3$ and $i, \,j, \,k, \,\dots = 1 - 3$,
respectively.

We define the metric compatible covariant derivative ${\bf \nabla}$
and so the spacetime connection $\Gamma^{a}{}_{bc}$ by
\be
\nabla_{a}V^{b} := \p_{a}(V^{b}) + \Gamma^{b}{}_{ca}\,V^{c} \ ,
\hsp5 \nabla_{a}V_{b} := \p_{a}(V_{b}) - \Gamma^{c}{}_{ba}\,V_{c}
\ .
\ee
In an orthonormal frame $\{\,\p_{a}\,\}$ we have
\be
0 = \nabla_{a}g_{bc} = \nabla_{a}\eta_{bc} = -\,\Gamma^{d}{}_{ba}\,
\eta_{dc} - \Gamma^{d}{}_{ca}\,\eta_{bd} \ ,
\ee
i.e., $\Gamma_{(ab)c} = 0$. For easier direct comparison it should
be noted that Friedrich's notation for the connection components is
$\nabla_{a}V^{b} := \p_{a}(V^{b}) + \Gamma_{a}{}^{b}{}_{c}\,V^{c}$
instead \ct{fri96,fri98} , such that the index referring to the
`left-acting' derivative is the first one to appear on
$\Gamma_{a}{}^{b}{}_{c}$. Compared to Ref. \ct{hveugg97}, we will
choose the fluid vorticity vector $\om^{a}$ to have the {\em
opposite\/} sign; thus, its definition will now comply with the
Newtonian convention. Finally we should remark that the Weyl
curvature variable $B_{ab}$ used by Friedrich \ct{fri96,fri98} has
the {\em opposite\/} sign to our $H_{ab}$, as in his definition the
negative of the 3-volume element was used. Our convention ensures
the {\em same\/} relative sign structure between $E_{ab}$ and
$H_{ab}$ in the Bianchi field equations when compared to that
between $E_{a}$ and $H_{a}$ in the Maxwell field equations.

\subsection{Evolution of Weyl curvature characteristic eigenfields}
\l{app2}

\noindent
{\em Transverse $v = \pm\,1$ modes\/}: \nopagebreak
\bea
(\p_{0}\mp\p_{1})\,(E_{-}\mp H_{1}) \pm \sfrac{1}{2}\,\p_{2}
(E_{3}\mp H_{2}) \mp \sfrac{1}{2}\,\p_{3}(E_{2}\pm H_{3})
& = & -\,(\Th-\sig_{+}\mp 2\,\udot_{1}\pm a_{1})\,(E_{-}\mp H_{1})
\nonumber \\
& & \hsp5 - \ (\om_{1}+2\,\Omega_{1}\pm n_{22}\pm n_{33}
\mp\sfrac{1}{2}\,n_{11})\,(E_{1}\pm H_{-})
\nonumber \\
& & \hsp5 - \ \sfrac{1}{2}\,(\sqrt{3}\,\sig_{2}-\om_{2}
-2\,\Omega_{2}\mp 2\,\udot_{3}\pm a_{3}\pm 3\,n_{12})\,(E_{2}\pm
H_{3}) \nonumber \\
& & \hsp5 + \ \sfrac{1}{2}\,(\sqrt{3}\,\sig_{3}+\om_{3}
+2\,\Omega_{3}\mp 2\,\udot_{2}\pm a_{2}\mp 3\,n_{31})\,(E_{3}\mp
H_{2}) \nonumber \\
& & \hsp5 - \ \sfrac{1}{2}\,(\mu+p)\,\sig_{-}
+ (\sig_{-}\mp \sqrt{3}\,n_{23})\,E_{+}
\nonumber \\
& & \hsp5 \mp\ (\sig_{1}\pm\sfrac{\sqrt{3}}{2}\,n_{22}
\mp\sfrac{\sqrt{3}}{2}\,n_{33})\,H_{+}
\\ \nonumber \\
(\p_{0}\mp\p_{1})\,(E_{1}\pm H_{-}) \pm \sfrac{1}{2}\,\p_{2}
(E_{2}\pm H_{3}) \pm \sfrac{1}{2}\,\p_{3}(E_{3}\mp H_{2})
& = & -\,(\Th-\sig_{+}\mp 2\,\udot_{1}\pm a_{1})\,(E_{1}\pm H_{-})
\nonumber \\
& & \hsp5 + \ (\om_{1}+2\,\Omega_{1}\pm n_{22}\pm n_{33}
\mp\sfrac{1}{2}\,n_{11})\,(E_{-}\mp H_{1})
\nonumber \\
& & \hsp5 + \ \sfrac{1}{2}\,(\sqrt{3}\,\sig_{2}-\om_{2}
-2\,\Omega_{2}\mp 2\,\udot_{3}\pm a_{3}\pm 3\,n_{12})\,(E_{3}\mp
H_{2}) \nonumber \\
& & \hsp5 + \ \sfrac{1}{2}\,(\sqrt{3}\,\sig_{3}+\om_{3}
+2\,\Omega_{3}\mp 2\,\udot_{2}\pm a_{2}\mp 3\,n_{31})\,(E_{2}\pm
H_{3}) \nonumber \\
& & \hsp5 - \ \sfrac{1}{2}\,(\mu+p)\,\sig_{1}
+ (\sig_{1}\pm\sfrac{\sqrt{3}}{2}\,n_{22}
\mp\sfrac{\sqrt{3}}{2}\,n_{33})\,E_{+}
\nonumber \\
& & \hsp5 \pm\ (\sig_{-}\mp \sqrt{3}\,n_{23})\,H_{+} \ .
\eea
Note that, with the exception of the source terms proportional to
$\sig_{-}$, $\sig_{1}$, $E_{+}$ and $H_{+}$, the {\em outgoing\/}
characteristic eigenfields are driven only by {\em outgoing\/}
characteristic eigenfields and, similarly, the {\em incoming\/}
characteristic eigenfields are driven only by {\em incoming\/}
characteristic eigenfields. \enl

\noindent
{\em Longitudinal $v = \pm\,\sfrac{1}{2}$ modes\/}: \nopagebreak
\bea
& & (\p_{0}\mp\sfrac{1}{2}\,\p_{1})\,(E_{3}\mp H_{2})
\pm \sfrac{1}{2}\,\p_{2}(E_{-}\mp H_{1}+\sqrt{3}\,E_{+})
\pm \sfrac{1}{2}\,\p_{3}(E_{1}\pm H_{-}\pm\sqrt{3}\,H_{+})
\nonumber \\
& & \hspace{35mm} = -\,\sfrac{1}{2}\,(2\,\Th+\sig_{+}\mp
2\,\udot_{1}\pm a_{1})\,(E_{3}\mp H_{2}) \nonumber \\
& & \hspace{40mm} + \ \sfrac{\sqrt{3}}{2}\,(\sig_{-}\mp\sqrt{3}\,
n_{23})\,(E_{3}\pm H_{2}) - \sfrac{1}{2}\,(\om_{1}+2\,\Omega_{1})\,
(E_{2}\pm H_{3}) + \sfrac{\sqrt{3}}{2}\,\sig_{1}\,(E_{2}\mp H_{3})
\nonumber \\
& & \hspace{40mm} + \ \sfrac{1}{2}\,(\sqrt{3}\,\sig_{3}-\om_{3}
-2\,\Omega_{3}\mp 2\,\udot_{2}\pm a_{2}\pm 3\,n_{31})\,(E_{-}\mp
H_{1}) \nonumber \\
& & \hspace{40mm} + \ \sfrac{1}{2}\,(\sqrt{3}\,\sig_{2}+\om_{2}
+2\,\Omega_{2}\mp 2\,\udot_{3}\pm a_{3}\mp 3\,n_{12})\,(E_{1}\pm
H_{-}) \nonumber \\
& & \hspace{40mm} - \ \sfrac{1}{2}\,(\mu+p)\,\sig_{3}
- \sfrac{\sqrt{3}}{2}\,(\sfrac{1}{\sqrt{3}}\,\sig_{3}+\om_{3}
+2\,\Omega_{3}\mp 2\,\udot_{2}\pm a_{2}\mp n_{31})\,E_{+}
\nonumber \\
& & \hspace{40mm} \pm\ \sfrac{\sqrt{3}}{2}\,(\sfrac{1}{\sqrt{3}}\,
\sig_{2}-\om_{2}-2\,\Omega_{2}\mp 2\,\udot_{3}\pm a_{3}\pm n_{12})
\,H_{+} \nonumber \\
& & \hspace{40mm} \mp\ \sfrac{1}{2}\,(2\,n_{33}+2\,n_{11}-n_{22})
\,E_{2} - \sfrac{1}{2}\,(2\,n_{11}+2\,n_{22}-n_{33})\,H_{3}
\\ \nonumber \\
& & (\p_{0}\mp\sfrac{1}{2}\,\p_{1})\,(E_{2}\pm H_{3})
\pm \sfrac{1}{2}\,\p_{2}(E_{1}\pm H_{-}\mp\sqrt{3}\,H_{+})
\mp \sfrac{1}{2}\,\p_{3}(E_{-}\mp H_{1}+\sqrt{3}\,E_{+})
\nonumber \\
& & \hspace{35mm} = -\,\sfrac{1}{2}\,(2\,\Th+\sig_{+}\mp
2\,\udot_{1}\pm a_{1})\,(E_{2}\pm H_{3}) \nonumber \\
& & \hspace{40mm} - \ \sfrac{\sqrt{3}}{2}\,(\sig_{-}\mp\sqrt{3}\,
n_{23})\,(E_{2}\mp H_{3}) + \sfrac{1}{2}\,(\om_{1}+2\,\Omega_{1})\,
(E_{3}\mp H_{2}) + \sfrac{\sqrt{3}}{2}\,\sig_{1}\,(E_{3}\pm H_{2})
\nonumber \\
& & \hspace{40mm} - \ \sfrac{1}{2}\,(\sqrt{3}\,\sig_{2}+\om_{2}
+2\,\Omega_{2}\mp 2\,\udot_{3}\pm a_{3}\mp 3\,n_{12})\,(E_{-}\mp
H_{1}) \nonumber \\
& & \hspace{40mm} + \ \sfrac{1}{2}\,(\sqrt{3}\,\sig_{3}-\om_{3}
-2\,\Omega_{3}\mp 2\,\udot_{2}\pm a_{2}\pm 3\,n_{31})\,(E_{1}\pm
H_{-}) \nonumber \\
& & \hspace{40mm} - \ \sfrac{1}{2}\,(\mu+p)\,\sig_{2}
- \sfrac{\sqrt{3}}{2}\,(\sfrac{1}{\sqrt{3}}\,\sig_{2}-\om_{2}
-2\,\Omega_{2}\mp 2\,\udot_{3}\pm a_{3}\pm n_{12})\,E_{+}
\nonumber \\
& & \hspace{40mm} \mp\ \sfrac{\sqrt{3}}{2}\,(\sfrac{1}{\sqrt{3}}\,
\sig_{3}+\om_{3}+2\,\Omega_{3}\mp 2\,\udot_{2}\pm a_{2}\mp n_{31})
\,H_{+} \nonumber \\
& & \hspace{40mm} \pm\ \sfrac{1}{2}\,(2\,n_{11}+2\,n_{22}-n_{33})
\,E_{3} - \sfrac{1}{2}\,(2\,n_{33}+2\,n_{11}-n_{22})\,H_{2} \ .
\eea
%

\subsection{Lorentz-invariant Weyl curvature scalars}
\l{app3}
In terms of the tracefree-adapted frame variables of Eq. (\r{apm})
and the characteristic eigenfields of the Weyl curvature, the four
standard Lorentz-invariant scalars assume the forms
\bea
\sfrac{1}{8}\,C^{ab}{}_{cd}\,C^{cd}{}_{ab} & = & E^{a}\!_{b}\,
E^{b}\!_{a} - H^{a}\!_{b}\,H^{b}\!_{a} \nonumber \\
& = & \sfrac{2}{3}\,[\ (E_{+})^{2} + (H_{+})^{2}
+ (E_{-}-H_{1})\,(E_{-}+H_{1}) + (E_{1}+H_{-})\,(E_{1}-H_{-})
\nonumber \\
& & \hspace{15mm} + \ (E_{3}-H_{2})\,(E_{3}+H_{2})
+ (E_{2}+H_{3})\,(E_{2}-H_{3})\ ]
\\ \nonumber \\
\sfrac{1}{8}\,{*}C^{ab}{}_{cd}\,C^{cd}{}_{ab} & = &
-\,2\,E^{a}\!_{b}\,H^{b}\!_{a} \nonumber \\
& = & -\,\sfrac{2}{3}\,[\ 2\,E_{+}\,H_{+}
- (E_{-}-H_{1})\,(E_{1}-H_{-}) + (E_{-}+H_{1})\,(E_{1}+H_{-})
\nonumber \\
& & \hspace{15mm} - \ (E_{3}-H_{2})\,(E_{2}-H_{3})
+ (E_{3}+H_{2})\,(E_{2}+H_{3})\ ]
\\ \nonumber \\
\sfrac{1}{16}\,C^{ab}{}_{cd}\,C^{cd}{}_{ef}\,C^{ef}{}_{ab}
& = & -\,E^{a}\!_{b}\,E^{b}\!_{c}\,E^{c}\!_{a}
+ 3\,E^{a}\!_{b}\,H^{b}\!_{c}\,H^{c}\!_{a} \nonumber \\
& = & \sfrac{2}{3}\,[\ \sfrac{1}{3}\,(E_{+})^{3} - E_{+}\,
(H_{+})^{2} - E_{+}\,(E_{-}-H_{1})\,(E_{-}+H_{1})
- E_{+}\,(E_{1}+H_{-})\,(E_{1}-H_{-}) \nonumber \\
& & \hspace{15mm} + \ \sfrac{1}{2}\,E_{+}\,(E_{3}-H_{2})\,
(E_{3}+H_{2}) + \sfrac{1}{2}\,E_{+}\,(E_{2}+H_{3})\,
(E_{2}-H_{3}) \nonumber \\
& & \hspace{15mm} - \ H_{+}\,(E_{-}-H_{1})\,(E_{1}-H_{-})
+ H_{+}\,(E_{-}+H_{1})\,(E_{1}+H_{-}) \nonumber \\
& & \hspace{15mm} + \ \sfrac{1}{2}\,H_{+}\,(E_{3}-H_{2})\,
(E_{2}-H_{3}) - \sfrac{1}{2}\,H_{+}\,(E_{3}+H_{2})\,
(E_{2}+H_{3}) \nonumber \\
& & \hspace{15mm} - \ \sfrac{\sqrt{3}}{4}\,(E_{-}-H_{1})\,
(E_{3}+H_{2})^{2} - \sfrac{\sqrt{3}}{4}\,(E_{-}+H_{1})\,
(E_{3}-H_{2})^{2} \nonumber \\
& & \hspace{15mm} + \ \sfrac{\sqrt{3}}{4}\,(E_{-}-H_{1})\,
(E_{2}-H_{3})^{2} + \sfrac{\sqrt{3}}{4}\,(E_{-}+H_{1})\,
(E_{2}+H_{3})^{2} \nonumber \\
& & \hspace{15mm} - \ \sfrac{\sqrt{3}}{2}\,(E_{1}+H_{-})\,
(E_{3}+H_{2})\,(E_{2}-H_{3}) \nonumber \\
& & \hspace{15mm} - \ \sfrac{\sqrt{3}}{2}\,(E_{1}-H_{-})\,
(E_{3}-H_{2})\,(E_{2}+H_{3})\ ]
\\ \nonumber \\
\sfrac{1}{16}\,{*}C^{ab}{}_{cd}\,{*}C^{cd}{}_{ef}\,{*}C^{ef}{}_{ab}
& = & H^{a}\!_{b}\,H^{b}\!_{c}\,H^{c}\!_{a}
- 3\,H^{a}\!_{b}\,E^{b}\!_{c}\,E^{c}\!_{a} \nonumber \\
& = & -\,\sfrac{2}{3}\,[\ \sfrac{1}{3}\,(H_{+})^{3} + H_{+}\,
(E_{+})^{2} + H_{+}\,(E_{-}-H_{1})\,(E_{-}+H_{1})
+ H_{+}\,(E_{1}+H_{-})\,(E_{1}-H_{-}) \nonumber \\
& & \hspace{15mm} - \ \sfrac{1}{2}\,H_{+}\,(E_{3}-H_{2})\,
(E_{3}+H_{2}) - \sfrac{1}{2}\,H_{+}\,(E_{2}+H_{3})\,
(E_{2}-H_{3}) \nonumber \\
& & \hspace{15mm} - \ E_{+}\,(E_{-}-H_{1})\,(E_{1}-H_{-})
+ E_{+}\,(E_{-}+H_{1})\,(E_{1}+H_{-}) \nonumber \\
& & \hspace{15mm} + \ \sfrac{1}{2}\,E_{+}\,(E_{3}-H_{2})\,
(E_{2}-H_{3}) - \sfrac{1}{2}\,E_{+}\,(E_{3}+H_{2})\,
(E_{2}+H_{3}) \nonumber \\
& & \hspace{15mm} + \ \sfrac{\sqrt{3}}{4}\,(E_{1}+H_{-})\,
(E_{3}+H_{2})^{2} - \sfrac{\sqrt{3}}{4}\,(E_{1}-H_{-})\,
(E_{3}-H_{2})^{2} \nonumber \\
& & \hspace{15mm} - \ \sfrac{\sqrt{3}}{4}\,(E_{1}+H_{-})\,
(E_{2}-H_{3})^{2} + \sfrac{\sqrt{3}}{4}\,(E_{1}-H_{-})\,
(E_{2}+H_{3})^{2} \nonumber \\
& & \hspace{15mm} - \ \sfrac{\sqrt{3}}{2}\,(E_{-}-H_{1})\,
(E_{3}+H_{2})\,(E_{2}-H_{3}) \nonumber \\
& & \hspace{15mm} + \ \sfrac{\sqrt{3}}{2}\,(E_{-}+H_{1})\,
(E_{3}-H_{2})\,(E_{2}+H_{3})\ ] \ .
\eea
%

\subsection{Weyl curvature variables in linear perturbation
formalisms of FLRW}
\l{app4}
Linearised perturbations to FLRW cosmological models are often
analysed in terms of the non-geometrical gauge-invariant variables
suggested by Bardeen \ct{bar80}. From the paper by Goode \ct{goo89}
(\,see also Ref. \ct{bruetal92}\,) it is straightforward to expand
the characteristic eigenfields of the Weyl curvature in terms of
the former. In the absence of dissipative perturbations to the
perfect fluid matter content it follows that $\Phi_{A} =
-\,\Phi_{H}$. \enl

\noindent
{\em Scalar modes\/}: \nopagebreak
\bea
\begin{array}{lcl}
E_{+} & = & -\,(3/4)\,k^{2}\,(\Phi_{A}-\Phi_{H})\,
Q^{(0)}{}_{11} \\
E_{-} & = & (\sqrt{3}/4)\,k^{2}\,(\Phi_{A}-\Phi_{H})\,
[\ Q^{(0)}{}_{22}-Q^{(0)}{}_{33}\ ] \\
E_{1} & = & (\sqrt{3}/4)\,k^{2}\,(\Phi_{A}-\Phi_{H})\,
Q^{(0)}{}_{23} \\
E_{2} & = & (\sqrt{3}/4)\,k^{2}\,(\Phi_{A}-\Phi_{H})\,
Q^{(0)}{}_{31} \\
E_{3} & = & (\sqrt{3}/4)\,k^{2}\,(\Phi_{A}-\Phi_{H})\,
Q^{(0)}{}_{12} \\
H_{\alpha\beta} & = & 0 \ .
\end{array}
\eea
{\em Vector modes\/}: \nopagebreak
\bea
\begin{array}{lcl}
E_{+} & = & -\,(3/4)\,k\,\dot{\Psi}\,Q^{(1)}{}_{11} \\
H_{+} & = & -\,(3/4)\,\Psi\,[\ Q^{(1)}{}_{2|13}
- Q^{(1)}{}_{3|12}\ ] \\
(E_{-}\mp H_{1}) & = & (\sqrt{3}/4)\,k\,\dot{\Psi}\,
[\ Q^{(1)}{}_{22}-Q^{(1)}{}_{33}\ ] \mp (\sqrt{3}/4)\,\Psi\,
[\ Q^{(1)}{}_{1|22} - Q^{(1)}{}_{1|33} - Q^{(1)}{}_{2|21}
+ Q^{(1)}{}_{3|31}\ ] \\
(E_{1}\pm H_{-}) & = & (\sqrt{3}/2)\,k\,\dot{\Psi}\,
Q^{(1)}{}_{23} \pm (\sqrt{3}/4)\,\Psi\,
[\ Q^{(1)}{}_{2|31} + Q^{(1)}{}_{3|21} - Q^{(1)}{}_{1|23}
- Q^{(1)}{}_{1|32}\ ] \\
(E_{3}\mp H_{2}) & = & (\sqrt{3}/2)\,k\,\dot{\Psi}\,
Q^{(1)}{}_{12} \mp (\sqrt{3}/4)\,\Psi\,
[\ Q^{(1)}{}_{2|33} - Q^{(1)}{}_{2|11} - Q^{(1)}{}_{3|32}
+ Q^{(1)}{}_{1|12}\ ] \\
(E_{2}\pm H_{3}) & = & (\sqrt{3}/2)\,k\,\dot{\Psi}\,
Q^{(1)}{}_{31} \pm (\sqrt{3}/4)\,\Psi\,
[\ Q^{(1)}{}_{3|11} - Q^{(1)}{}_{3|22} - Q^{(1)}{}_{1|13}
+ Q^{(1)}{}_{2|23}\ ] \ .
\end{array}
\eea
{\em Tensor modes\/}: \nopagebreak
\bea
\begin{array}{lcl}
E_{+} & = & (3/4)\,[\ \ddot{H}{}_{T}^{(2)} - (k^{2}+2K)\,
H_{T}^{(2)}\ ]\,Q^{(2)}{}_{11} \\
H_{+} & = & -\,(3/2)\,\dot{H}{}_{T}^{(2)}\,[\ Q^{(2)}{}_{12|3}
- Q^{(2)}{}_{31|2}\ ] \\
(E_{-}\mp H_{1}) & = & -\,(\sqrt{3}/4)\,[\ \ddot{H}{}_{T}^{(2)}
- (k^{2}+2K)\,H_{T}^{(2)}\ ]\,[\ Q^{(2)}{}_{22}-Q^{(2)}{}_{33}\ ]
\\
& & \hsp5 \pm \,(\sqrt{3}/2)\,\dot{H}{}_{T}^{(2)}\,
[\ Q^{(2)}{}_{22|1} - Q^{(2)}{}_{33|1} + Q^{(2)}{}_{31|3}
- Q^{(2)}{}_{12|2}\ ] \\
(E_{1}\pm H_{-}) & = & -\,(\sqrt{3}/4)\,[\ \ddot{H}{}_{T}^{(2)}
- (k^{2}+2K)\,H_{T}^{(2)}\ ]\,Q^{(2)}{}_{23} \\
& & \hsp5 \pm \,(\sqrt{3}/2)\,\dot{H}{}_{T}^{(2)}\,
[\ 2\,Q^{(2)}{}_{23|1} - Q^{(2)}{}_{31|2} - Q^{(2)}{}_{12|3}\ ] \\
(E_{3}\mp H_{2}) & = & -\,(\sqrt{3}/4)\,[\ \ddot{H}{}_{T}^{(2)}
- (k^{2}+2K)\,H_{T}^{(2)}\ ]\,Q^{(2)}{}_{12} \\
& & \hsp5 \pm \,(\sqrt{3}/2)\,\dot{H}{}_{T}^{(2)}\,
[\ Q^{(2)}{}_{33|2} - Q^{(2)}{}_{11|2} + Q^{(2)}{}_{12|1}
- Q^{(2)}{}_{23|3}\ ] \\
(E_{2}\pm H_{3}) & = & -\,(\sqrt{3}/4)\,[\ \ddot{H}{}_{T}^{(2)}
- (k^{2}+2K)\,H_{T}^{(2)}\ ]\,Q^{(2)}{}_{31} \\
& & \hsp5 \mp \,(\sqrt{3}/2)\,\dot{H}{}_{T}^{(2)}\,
[\ Q^{(2)}{}_{11|3} - Q^{(2)}{}_{22|3} + Q^{(2)}{}_{23|2}
- Q^{(2)}{}_{31|1}\ ] \ .
\end{array}
\eea
%

\subsection{The Weyl curvature force term in the deviation equation
for null and timelike congruences}
\l{app5}
We choose the tangents to the {\em null\/} curves to be given by
$k^{a} := E\,(u^{a}+e^{a})$, where $E := -(u_{a}\eta^{a})$. Then
the orthogonal deviation vector can be expressed as $\eta^{a} =
-\,E^{-1}\,(u_{b}\eta^{b}) \,k^{a} + \eta_{2}\,e_{2}{}^{a} +
\eta_{3}\,e_{3}{}^{a}$. Thus, the force term is \enl

\noindent
{\em Null congruences\/}:
\bea
C^{a}{}_{bcd}\,k^{b}\,\eta^{c}\,k^{d} & = & \sfrac{1}{\sqrt{3}}\,
E\,[\ (E_{3}-H_{2})\,\eta_{2} + (E_{2}+H_{3})\,\eta_{3}\ ]\ k^{a}
+ \sfrac{2}{\sqrt{3}}\,E^{2}\,[\ (E_{-}-H_{1})\,\eta_{2}
+ (E_{1}+H_{-})\,\eta_{3}\ ]\ e_{2}{}^{a} \nonumber \\
& & \hsp5 + \ \sfrac{2}{\sqrt{3}}\,E^{2}\,[\ (E_{1}+H_{-})\,
\eta_{2} - (E_{-}-H_{1})\,\eta_{3}\ ]\ e_{3}{}^{a} \ .
\eea
For {\em timelike\/} curves moving with velocity $v^{a} :=
v\,e^{a}$ relative to ${\bf u}$ we define the unit tangents by
$V^{a} := \gam\, (u^{a}+v\,e^{a})$, where $\gam :=
(1-v^{2})^{-1/2}$. Then the orthogonal deviation vector can be
expressed as $\eta^{a} = (e_{b}\eta^{b})\,(v\,u^{a}+e^{a}) +
\eta_{2}\,e_{2}{}^{a} + \eta_{3}\,e_{3}{}^{a}$. Thus, the force
term is \enl

\noindent
{\em Timelike congruences\/}:
\bea
C^{a}{}_{bcd}\,V^{b}\,\eta^{c}\,V^{d} & = & -\,\sfrac{1}{\sqrt{3}}
\,[\ \sfrac{2}{\sqrt{3}}\,E_{+}\,\eta_{1} - \gam^{2}\,(E_{3}
-v\,H_{2})\,\eta_{2} - \gam^{2}\,(E_{2}+v\,H_{3})\,\eta_{3}\ ]
\ (v\,u^{a}+e^{a}) \nonumber \\
& & \hsp5 + \ \sfrac{1}{\sqrt{3}}\,[\ (E_{3}-v\,H_{2})\,\eta_{1}
+ \gam^{2}\,(\sfrac{1}{\sqrt{3}}\,E_{+}-\sfrac{1}{\sqrt{3}}\,v^{2}
\,E_{+}+E_{-}+v^{2}\,E_{-}-2\,v\,H_{1})\,\eta_{2} \nonumber \\
& & \hspace{2cm} + \ \gam^{2}\,(E_{1}+v^{2}\,E_{1}+2\,v\,H_{-})\,
\eta_{3}\ ]\ e_{2}{}^{a} \nonumber \\
& & \hsp5 + \ \sfrac{1}{\sqrt{3}}\,[\ (E_{2}+v\,H_{3})\,\eta_{1}
+ \gam^{2}\,(E_{1}+v^{2}\,E_{1}+2\,v\,H_{-})\,\eta_{2} \nonumber \\
& & \hspace{2cm} + \ \gam^{2}\,(\sfrac{1}{\sqrt{3}}\,E_{+}
-\sfrac{1}{\sqrt{3}}\,v^{2}\,E_{+}-E_{-}-v^{2}\,E_{-}+2\,v\,H_{1})
\,\eta_{3}\ ]\ e_{3}{}^{a} \ .
\eea
%



\end{document}